\documentclass[12pt]{iopart}
\usepackage[pdftex]{graphicx}

\usepackage{amssymb}
\usepackage{iopams}
\usepackage{setstack}

\begin{document}

\title{Two-photon spectra of quantum emitters}

\author{Alejandro Gonzalez-Tudela}
\address{F{\'i}sica Te{\'o}rica de la Materia Condensada, Universidad
  Aut{\'o}noma de Madrid, 28049, Madrid, Spain}

\author{Fabrice P. Laussy}
\address{F{\'i}sica Te{\'o}rica de la Materia Condensada, Universidad
  Aut{\'o}noma de Madrid, 28049, Madrid, Spain}

\author{Carlos Tejedor}
\address{F{\'i}sica Te{\'o}rica de la Materia Condensada, Universidad
  Aut{\'o}noma de Madrid, 28049, Madrid, Spain}

\author{Michael J. Hartmann}
\address{Physik Department, Technische Universit{\"a}t M{\"u}nchen,
  James Franck Str., 85748, Garching, Germany}

\author{Elena del Valle}
\address{Physik Department, Technische Universit{\"a}t M{\"u}nchen,
  James Franck Str., 85748, Garching, Germany}
\ead{elena.delvalle.reboul@gmail.com}

\begin{abstract}
  We apply our recently developed theory of frequency-filtered and
  time-resolved $N$-photon correlations~\cite{delvalle12a} to study
  the two-photon spectra of a variety of systems of increasing
  complexity: single mode emitters with two limiting statistics (one
  harmonic oscillator or a two-level system) and the various
  combinations that arise from their coupling. We consider both the
  linear and nonlinear regimes under incoherent excitation. We find
  that even the simplest systems display a rich dynamics of emission,
  not accessible by simple single photon spectroscopy.  In the strong
  coupling regime, novel two-photon emission processes involving
  virtual states are revealed. Furthermore, two general results are
  unraveled by two-photon correlations with narrow linewidth
  detectors: $i)$ filtering induced bunching and $ii)$ breakdown of
  the semi-classical theory. We show how to overcome this shortcoming
  in a fully-quantized picture.
\end{abstract}

% FPL shortcuts
\newcommand{\ud}[1]{#1^{\dagger}} 
\newcommand{\bra}[1]{\left\langle #1\right|}
\newcommand{\ket}[1]{\left| #1\right\rangle}
\newcommand{\braket}[2]{\langle #1|#2\rangle}
\newcommand{\mean}[1]{\langle#1\rangle}

\maketitle

\section{Introduction}
\label{intro}

At the heart of the modern theory of light lies the concept of the
photon, the quantum of the electromagnetic
field~\cite{dirac30_booka}. Regardless of the wave behaviour of light
in a variety of contexts, ultimately, every measurement is accounted
for by clicks on a detector. When the number of quanta is so large
that the granularity of the field disappears due to coarse-graining,
the limit of classical electrodynamics is
recovered~\cite{jackson_book75a}. It is one of the wonders of quantum
mechanics that the light field still exhibits wave-like properties
when tracked at the single-particle level. Most popularly, the
interference patterns in a double slit experiment is reconstructed by
the accumulation of a large number of single clicks, each observed in
isolation~\cite{tonomura89a}.

Usually, even in the regime of a small number of quanta in the field,
the experimental observables are obtained by averaging over a large
enough repetition of the experiment. For instance, a luminescence
spectrum is obtained by integrating the signal over long times, such
as to obtain a continuous distribution that describes the probability
of emission of each photon at a given energy. An example for such a
one-photon spectrum (1PS) $S(\omega)$ is given in Fig.~\ref{fig:1}(a)
for the simplest conceivable case, that of a single mode emitter. The
lineshape is simply a Lorentzian with a linewidth that is given by the
inverse lifetime, corresponding, in the time domain, to an exponential
decay of the excitation: the emission is memoryless and has the same
probability to occur at any given time, as befits a quantum
particle. This results in a spread of energy of the emitted photon.
Arguably, however, this lineshape is also typical of a classical
system, in the form of a damped resonance.

To access the dynamics of photons at a quantum level, with no
counterpart from a classical theory involving a continuous field, it
is necessary to invoke $N$-photon correlations. An average is still
taken over a large number of single-detection events, but the latter
now involves $N$ correlated clicks in each shot. In this text, we will
focus on the simplest case where $N=2$. There is an ample literature
on two-photon correlations~\cite{glauber06a}. Typically, photons are
detected as a function of time (and/or other commuting variables like
space) in photon counting configurations such as the celebrated
Hanburry Brown--Twiss setup~\cite{hanburybrown56a}.  The specificity
of our approach is that we study correlations in the frequency domain,
i.e., we compute correlations between a photon detected at an
energy~$\omega_1$ and another at energy $\omega_2$. The conjugate time
information is still present through the linewidth of the frequency
windows. Real time can also be included, computing correlations
between a photon of frequency~$\omega_1$ detected at time~$t_1$ and
another of frequency~$\omega_2$ detected at time~$t_2$, although we
will limit here to simultaneous detection for brevity.  This is a
generalisation of the conventional (color-blind) theory. The
two-photon spectroscopy of a quantum emitter is obtained when such
correlations are computed (theoretically) or measured (experimentally)
for all the energies spanning the range of emission of the
system. Experimentally, the implementation is straightforward, as it
essentially consists in inserting filters between the system and the
conventional photon-counting setup. The technique is increasingly
popular in the laboratory~\cite{akopian06a, hennessy07a, kaniber08a,
  sallen10a, ulhaq12a} although its interpretation relies chiefly on
physical grounds or numerical simulations.  Theoretically, almost all
the effort has been dedicated to resonance
fluorescence~\cite{cohentannoudji79a, dalibard83a, knoll86a,
  nienhuis93a, bel09a}, owing to the difficulty of actual computation
for other systems, prior to our input~\cite{delvalle12a} that yields
exact results for arbitrary systems.

In this text, we present a comprehensive analysis of a variety of
fundamental quantum optical systems and show the extremely rich
dynamics and the novel insights that such frequency-resolved
two-photon correlations unravel. In
Section~\ref{eq:ThuJun14160425CEST2012}, we give a brief overview of
our theory to compute two-photon spectra (2PS), denoted
$g_\Gamma^{(2)}(\omega_1,\omega_2)$. The notation refers to the
Glauber precept that places photon correlations as the pillar of
quantum optics~\cite{glauber06a}.  The mode which photons are
collected from will be appended in square brackets for clarity but
only when necessary to avoid confusion.  The full theory for
$N$-photon correlations and/or arbitrary time delays is laid down in
Ref.~\cite{delvalle12a}. In Section~\ref{singlemode}, we address the
simplest case of a single mode: an harmonic oscillator or a two-level
system. These serve as the first illustration of our theme that the
2PS reveals a considerably richer picture of the dynamics of emission
for an otherwise identical 1PS lineshape, depending on both the mode
statistics or the nature of its broadening (homogeneous or
inhomogeneous). In Section~\ref{doublemode}, we analyse the more
complex dynamics of coupled modes. Here again, even though the
low-pumping regime results in identical 1PS lineshapes, the coupling
of two harmonic oscillators, two two-level systems, or one harmonic
oscillator to a two-level system (Jaynes-Cummings model), show
distinct and peculiar features in the 2PS, unravelling the different
nature of the three couplings. In the latter cases, of great
importance in cavity-QED, we thereby identify an entire class of
two-photon emission processes, to which one is completely oblivious
through one-photon spectroscopy. In
Section~\ref{sec:ThuJun14164126CEST2012}, we address the high
excitation regime of a two-level system, considering two ways of
driving it coherently: with the standard semi-classical
model~\cite{mollow69a}, that is, through the coupling to a classical
term in the Hamiltonian, describing an ideal laser, and with a
fully-quantized model~\cite{delvalle10d} where a very large number of
photons builds up a coherent field in a cavity. Both provide the
expected Mollow triplet lineshapes in the 1PS but only the full
quantum model provides a physical 2PS. In
Section~\ref{sec:ThuJun14165142CEST2012}, we conclude and discuss
further directions.

\section{Theory of one- and two-photon spectra}
\label{eq:ThuJun14160425CEST2012}

We have recently shown how to compute efficiently and easily
two-photon correlations between photons of frequencies $\omega_1$ and
$\omega_2$ through the coupling of two ``sensors'' to the open quantum
system of interest~\cite{delvalle12a}.  In the limit of vanishing
coupling of the sensors to the mode which photons correlations are to
be measured, the sensors intensity-intensity cross-correlations
recover exactly the frequency-resolved two-photon correlations defined
in the literature through integrals that are, however, too cumbersome
to compute directly in most cases~\cite{vogel_book06a}.  In the case
of zero delay between the photons, the normalised correlation
function, or \emph{two-photon spectrum} (2PS), is therefore obtained
simply in terms of the population operators of the sensors $n_1$ and
$n_2$, as:
\begin{equation}
  \label{eq:twophotonspectrum}
  g_{\Gamma_1,\Gamma_2}^{(2)}(\omega_1,\omega_2)=\lim_{\varepsilon \rightarrow 0}\frac{\mean{n_1 n_2}}{\mean{n_1}\mean{n_2}}\,,
\end{equation}
where $\varepsilon$ represents the coupling to the sensors, small
enough not to disturb the system, and $\Gamma_1$, $\Gamma_2$ are the
sensors decay rates that provide the detectors linewidths or, in other
terms, the respective widths of the frequency windows.  The case of
equal frequencies $g^{(2)}_\Gamma(\omega,\omega)$ measures the
photon-statistics of photons detected in a frequency window of
Lorentzian shape of linewidth $\Gamma$. This corresponds to applying a
single filter or to the effect of the detectors resolution in a
colour-blind measurement, and is thus also of prime interest.

In practical terms, the calculation can be performed in two
alternative ways, as discussed in Ref.~\cite{delvalle12a}. The first
method relies in the actual inclusion of the sensors in the dynamics
of the system, supplementing the Hamiltonian and the master equation
with the corresponding terms. The main drawback is that the size of
the Hilbert space is increased by a factor $2^2$ (and the density
matrix by a factor $4^2$). However, the computational operations are
simple. In one of the possible computational approaches to the
problem, one merely needs to solve the steady state of the full system
(a set of homogeneous linear equations), which only requires the
inversion of a matrix~\cite{delvalle09a}. An alternative method,
presented in detail in the supplemental material of
Ref.~\cite{delvalle12a}, consists in using the formal semi-analytical
expressions for the populations and cross correlations between the
sensors to leading order in the couplings, $\varepsilon^2$ and
$\varepsilon^4$ respectively. Then, the relevant quantities
$\mean{n_1}$, $\mean{n_2}$ and $\mean{n_1 n_2}$ can be obtained
through these expressions (that are provided below) and the master
equation of the system only with no increase in the numerical
complexity.  In order to implement this method, one first needs to
determine a vector of the system operators whose average are needed to
compute the coupling to the sensors. Calling $O$ the annihilation
operator of the mode of interest, which correlations one wishes to
compute, such a vector can be written in the form
$\mathbf{v}=(1,O,\ud{O},\ud{O}O\dots)$ (with subsequent terms
depending on the systems and its dynamics). The observables of
interest for the sensors cross-correlations are specified by the mean
values of this vector: $\langle\mathbf{v}\rangle =(1, \mean{O},
\mean{\ud{O}},\mean{\ud{O}O},\dots)$. A regression matrix~$\mathbf{M}$
can be obtained from the master equation of the
system~\cite{delvalle09a} which rules the dynamics of these
correlators according to $\partial_t\langle\mathbf{v}\rangle =
\mathbf{M}\langle\mathbf{v}\rangle$. The steady state (if it exists)
is obtained in the limit of infinite times:
$\langle\mathbf{v}^\mathrm{ss}\rangle=\lim_{t\rightarrow\infty}e^{\mathbf{M}
  t}\langle\mathbf{v}(0)\rangle$, for any initial state
$\mathbf{v}(0)$. Next, one builds two re-ordering matrices,
$\mathbf{T}_\pm$, which, when acting on the vector
$\langle\mathbf{v}\rangle$, introduce in all correlators an extra
operator $\ud{O}$ for $\mathbf{T}_+$ and an extra operator $O$ for
$\mathbf{T}_-$, keeping normal order in each case. That is,
$\mathbf{T}_+\langle\mathbf{v}\rangle
=(\mean{\ud{O}},\mean{\ud{O}O},\mean{(\ud{O})^2},\mean{(\ud{O})^2O},\dots)$
and $\mathbf{T}_-\langle\mathbf{v}\rangle
=(\mean{O},\mean{O^2},\mean{\ud{O}O},\mean{\ud{O}O^2},\dots)$,
respectively. With these matrices, one obtains the populations of the
sensors to leading order in $\varepsilon$, as the first element
$[\dots]_1$ of the vector:
\begin{equation}
  \label{eq:WedMar21032339CET2012}
%  \mean{n_i}=\frac{2 \varepsilon^2}{\Gamma_i}\Re\Big[\mathbf{T}_+ \frac{-1}{\mathbf{M}+[-i\omega_i-\frac{\Gamma_i}{2}]\mathbf{1}}\mathbf{T}_-\langle\mathbf{v}^\mathrm{ss}\rangle\Big]_1\,,\quad i=1,\,2\,.
  \mean{n_j}=\frac{2 \varepsilon^2}{\Gamma_j}\Re\Big[\mathbf{T}_+ \frac{-1}{\mathbf{M}+[-i\omega_j-\frac{\Gamma_j}{2}]\mathbf{1}}\mathbf{T}_-\langle\mathbf{v}^\mathrm{ss}\rangle\Big]_1\,,\quad j=1,\,2\,.
\end{equation}
Regarded as a function of the frequency $\omega_j$, this produces the
1PS as measured by a detector of linewidth~$\Gamma_j$. Similarly, the
cross-correlations that provide the 2PS is given~by:
\begin{eqnarray}
  \label{eq:WedMar21195020CET2012}
  \fl\mean{n_1 n_2}=\frac{2 \varepsilon^4}{\Gamma_1+\Gamma_2}\Re \Big[ \mathbf{T}_+ \frac{-1}{\mathbf{M}+(-i\omega_2-\Gamma_1-\frac{\Gamma_2}2)\mathbf{1}}\nonumber\\
  \fl\times \Big\{ \mathbf{T}_- \frac{-1}{\mathbf{M}-\Gamma_1\mathbf{1}}
  \Big( \mathbf{T}_+ \frac{-1}{\mathbf{M}+(-i\omega_1-\frac{\Gamma_1}2)\mathbf{1}}\mathbf{T}_- + \mathbf{T}_-
  \frac{-1}{\mathbf{M}+(i\omega_1-\frac{\Gamma_1}2)\mathbf{1}}  \mathbf{T}_+  \Big)\nonumber\\
  \fl+ \mathbf{T}_-
  \frac{-1}{\mathbf{M}+(i\omega_1-i\omega_2-\frac{\Gamma_1+\Gamma_2}2)\mathbf{1}}
  \Big( \mathbf{T}_- \frac{-1}{\mathbf{M}+(i\omega_1-\frac{\Gamma_1}2)\mathbf{1}}  \mathbf{T}_+ + \mathbf{T}_+ \frac{-1}{\mathbf{M}+(-i\omega_2-\frac{\Gamma_2}2)\mathbf{1}}\mathbf{T}_-  \Big)\nonumber\\
  \fl+\mathbf{T}_+\frac{-1}{\mathbf{M}+(-i\omega_1-i\omega_2-\frac{\Gamma_1+\Gamma_2}2)\mathbf{1}}
  \mathbf{T}_-
  \Big(\frac{-1}{\mathbf{M}+(-i\omega_2-\frac{\Gamma_2}2)\mathbf{1}} +\frac{-1}{\mathbf{M}+(-i\omega_1-\frac{\Gamma_1}2)\mathbf{1}} \Big)\mathbf{T}_- \Big\}\mathbf{v}^\mathrm{ss}\Big]_1\nonumber\\
  \fl+[1\leftrightarrow 2]\,,
\end{eqnarray}
where $\left[1\leftrightarrow 2\right]$ means the interchange of
sensors 1 and 2, that is, permuting $\omega_1\leftrightarrow \omega_2$
and $\Gamma_1\leftrightarrow \Gamma_2$ everywhere.

The advantage of this second method is twofold. First, it is very
useful when the Hilbert space is small as it can lead to closed-form
analytical expressions, as we show in the next Section. Second, it may
also be numerically advantageous as the matrices involved correspond
to the original (smaller) Hilbert space of the bare system, in the
absence of sensors. The computational price to pay is in a higher
number of matrix operations (eleven different matrix inversions and
numerous multiplications for $\mean{n_1 n_2}$). In some situations,
the advantage of a smaller Hilbert space dominates. In others, it is
more convenient and straightforward to explicitly include the sensors
and solve for the enlarged system, with a similar overall numerical
efficiency. We used both methods indistinctively to obtain the results
presented in the rest of this text.

From now on, the linewidths of the detectors will be taken equal,
$\Gamma_1=\Gamma_2=\Gamma$, for the sake of simplicity. We will thus
simply write $g_{\Gamma}^{(2)}(\omega_1,\omega_2)$. The width $\Gamma$
of the frequency windows is a fundamental parameter that cannot be
dispensed with, unlike the 1PS case where an ideal detector is usually
assumed in theoretical works. In the 2PS, $\Gamma$ provides the
uncertainty in the frequency ($\Gamma$) and time of detection
($1/\Gamma$)~\cite{eberly77a}, as required by the Heisenberg
principle. In the limit of very broad filters ($\Gamma\rightarrow
\infty$) the standard (color-blind) second order coherence function at
zero delay is recovered: $\lim_{\Gamma\rightarrow
  \infty}g_{\Gamma}^{(2)}(\omega_1,\omega_2)=g^{(2)}$. The opposite
limit of very narrow filters ($\Gamma\rightarrow 0$), results in a
systematic tendency regardless of the underlying system, namely
photons are uncorrelated, and those with the same frequency are
thermally bunched. This can be understood on physical grounds: in
order for the detectors to provide high precision in frequency
($\Gamma \rightarrow 0$), their interaction time with the system has
to be long ($1/\Gamma\rightarrow \infty$), so that the collected
photons correspond to all the possible times in the dynamics. This
leads to an apparent uncorrelated statistics: $\lim_{\Gamma\rightarrow
  0} g_{\Gamma}^{(2)}(\omega_1,\omega_2)=1$ provided that
$\omega_1\neq\omega_2$. The limiting value becomes
$\lim_{\Gamma\rightarrow 0} g_{\Gamma}^{(2)}(\omega,\omega)=2$ in the
case of detection at equal frequencies as there are $2!$ ways that two
indistinguishable photons can be collected by two different
detectors. We refer to this correlation tendency to bunch when
$\omega_1\approx \omega_2$ and $\Gamma$ is small (as compared to the
relevant linewidths in the system), as {\it indistinguishability
  bunching}. This effect has been observed experimentally in the
filtering of a single mode laser~\cite{centenoneelen93a}.

\section{Single mode emitters}
\label{singlemode}

Two-photon frequency-resolved spectroscopy is such a nascent field of
optical characterization that even the trivial systems need to be
investigated. Namely, our starting point is the free mode (or, in the
quantum optics terminology, ``single mode''), which Hamiltonian simply
reads:
\begin{equation}
  H_{O}=\omega_O\ud{O}O\,.
\end{equation}
There are two fundamental possibilities for the operator~$O$, namely,
it can be the annihilation operator of an harmonic oscillator, in
which case we denote it~$O=a$, or it can be that of a two-level
system, in which case we denote it~$O=\sigma$. Their respective
quantum algebra is given by the bosonic commutation rule
($a\ud{a}-\ud{a}a=1$) and the fermionic anticommutation rule
($\sigma\ud{\sigma}+\ud{\sigma}\sigma=1$).  The dynamics of the
density matrix $\rho$ of the emitter---including decay (required to
bridge with the external world where the measurement is performed) and
a continuous incoherent pump (to populate the system)---is given by
the master equation
\begin{equation}
  \label{eq:MonNov19183024CET2012}
  \partial_t\rho=i[\rho,H_O]+\frac{\gamma_O}2\mathcal{L}_{O}(\rho)+\frac{P_O}2\mathcal{L}_{\ud{O}}(\rho)\,,
\end{equation}
with the Lindblad super-operator $\mathcal{L}_{O}(\rho)=(2
O\rho\ud{O}-\ud{O}O\rho-\rho\ud{O}O)$.

\begin{figure}[t] 
  \centering
  \includegraphics[width=0.9\linewidth]{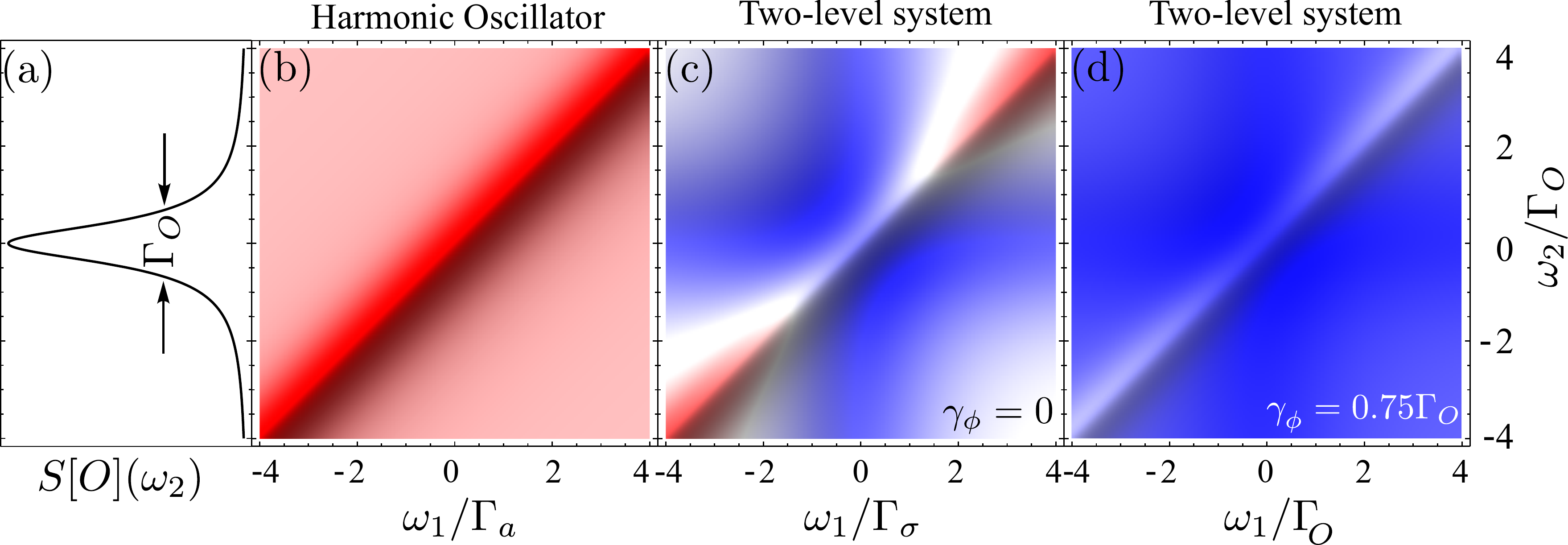}
 \caption{(a) The one-photon spectrum (1PS) common to a single
    harmonic oscillator and a two-level system, and the corresponding
    two-photon spectra (2PS),
    $g_{\Gamma_O}^{(2)}[O](\omega_1,\omega_2)$, for (b) an harmonic
    oscillator and (c)--(d) a two-level system, in units of the total
    broadening $\Gamma_O$. In (d), the emitter is subject to pure
    dephasing, $\gamma_\phi=0.75\Gamma_\sigma$, with $\Gamma_\sigma$
    such that the 1PS remains the one shown in (a). Beyond the
    bunching on the diagonal, common to all 2PS, clear deviations are
    observed depending on the type of emitter: the harmonic oscillator
    is structureless while the two-level system concentrates its
    antibunching in a butterfly-shaped region of non-overlapping
    frequencies. Dephasing favours antibunching. The color code is: $0$ blue; $1$ white; $2$ red.}
  \label{fig:1}
\end{figure}

The 1PS, provided by Eq.~(\ref{eq:WedMar21032339CET2012}), is the same
for both systems $O=a,\sigma$. It is plotted in
Fig.~\ref{fig:1}(a). It is a Lorentzian lineshape with linewidth
$\Gamma_O$ (inverse mode lifetime) given by $\Gamma_a=\gamma_a-P_a$
for the harmonic oscillator and $\Gamma_\sigma=\gamma_\sigma+P_\sigma$ for the
two-level system~\cite{delvalle_book10a}.

The two-photon correlations, however, are dramatically different,
since the harmonic oscillator displays bunching, $g^{(2)}[a]=2$, while
the two-level system displays the exact opposite behaviour,
antibunching, $g^{(2)}[\sigma]=0$ (the mode which statistics is
measured is here denoted explicitly in square brackets to lift the
ambiguity). The 2PS also exhibits structures in the frequency
correlations that are qualitatively different. For such simple cases,
Eq.~(\ref{eq:WedMar21195020CET2012}) can be solved exactly and
analytical formulas for the 2PS obtained:
\begin{equation}
  \label{eq:TueOct2143944CEST2012}
  g_{\Gamma}^{(2)}[O](\omega_1,\omega_2)=\frac{\Gamma_O}{(\Gamma+\Gamma_O)^2}\Big[\Gamma_O+\frac{\Gamma^2 (2\Gamma +\Gamma_O)}{\Gamma^2+(\omega_1-\omega_2)^2}+\tilde g_{\Gamma}^{(2)}[O](\omega_1,\omega_2)\Big]
\end{equation}
with a common expression regardless of the mode~$O=a,\sigma$ spelt out
above, and a term $\tilde g_{\Gamma}^{(2)}[O]$ specific to each case,
given by:
\begin{equation}
  \label{twophotonHO}  
  \tilde g_{\Gamma}^{(2)}[a]=2\Gamma\left(1+\frac{\Gamma}{\Gamma_a}\right)\,,
\end{equation}
for the harmonic oscillator (note that it is frequency independent), and by
\begin{equation}
  \label{twophotonLS}
  \tilde g_{\Gamma}^{(2)}[\sigma](\omega_1,\omega_2)=4\Gamma-2\Gamma(2\Gamma+\Gamma_\sigma)\left( \frac{\frac{ 3\Gamma+\Gamma_\sigma}{2}}{(\frac{3\Gamma+\Gamma_\sigma}{2})^2+\omega_1^2} +\frac{ \frac{3\Gamma+\Gamma_\sigma}{2}}{(\frac{3\Gamma+\Gamma_\sigma}{2})^2+\omega_2^2} \right)\nonumber\,,
\end{equation}
%
%They become more transparent in:
%
%\begin{equation}
%  \label{twophotonHOa}
%  \mathrm{HO:}\quad g_{\Gamma_a,a}^{(2)}(\omega_1,\omega_2)=\frac{5}{4}+\frac{3}{4}\frac{\Gamma_a^2}{\Gamma_a^2+(\omega_1-\omega_2)^2}\,,
%\end{equation}
%
%and
%
%\begin{equation}
%  \label{twophotonLSa}
%  \mathrm{TLS:}\quad g_{\Gamma_\sigma,\sigma}^{(2)}(\omega_1,\omega_2)=\frac{5}{4}+\frac{3}{4}\frac{\Gamma_\sigma^2}{\Gamma_\sigma^2+(\omega_1-\omega_2)^2}-3\Big(\frac{\Gamma_\sigma^2}{4\Gamma_\sigma^2+\omega_1^2}+\frac{\Gamma_\sigma^2}{4\Gamma_\sigma^2+\omega_2^2}\Big)\,,
%\end{equation}
%
%
for the two-level system. The single filter case
$g^{(2)}_\Gamma(\omega,\omega)$ shows that for the two-level system,
correlations are maximum when filtering the peak itself, $\omega=0$,
and that a large overlap of the peak is needed to recover a good
antibunching of the two-level system (the ideal result is recovered in
the limit of broadband detection:
$\lim_{\Gamma\rightarrow\infty}g_{\Gamma}^{(2)}[\sigma](\omega_1,\omega_2)=g^{(2)}[\sigma]=0$). The
photon-counting correlation of the filtered peak indeed reads
$g^{(2)}_\Gamma[\sigma](0,0)=2(\Gamma_\sigma/\Gamma)\big/\big(3+\Gamma_\sigma/\Gamma\big)$,
e.g., the filter linewidth must be over sixty-six times that of the
emitter to reach the 1\% level of accuracy, an awkward requirement in
practice. For this reason, post-processing of the coincidences through
deconvolution of the spectral response is important to characterise
fairly a quantum emitter.

For the harmonic oscillator, on the other hand,
$g_\Gamma^{(2)}[a](\omega,\omega)=2$ regardless of the filtering
window~$\Gamma$ and of which part of the peak is detected (maximum or
any point in the tail). This is a manifestation of the classical
character of the harmonic oscillator: it has no local information, any
part behaves like the whole and one cannot tell apart from a filtered
window whether the information is of a microscopic or macroscopic
nature. In contrast, the two-level system has an energy scale and the
information is localised in the energy window proper to the dynamics
of the emitter.

The full 2PS of these systems further reveals such fundamental aspects
of these emitters. In Fig.~\ref{fig:1}, we plot
Eq.~(\ref{eq:TueOct2143944CEST2012}) in the case where the detector
linewidth matches that of the 1PS peak, $\Gamma=\Gamma_O$. In this and
all subsequent density plots of the 2PS, we use a color code where red
corresponds to bunching ($g^{(2)}\gtrsim2$), blue to antibunching
($g^{(2)}\approx0$) and white to non-correlated emission
($g^{(2)}=1$). We are more concerned in this text with the qualitative
patterns that emerge in two-photon spectroscopy than quantitative
results (the optimisation of given correlations is straightforward but
would bring us to a too voluminous discussion).

The common characteristic between the harmonic oscillator and the
two-level system is the indistinguishability bunching line, for
frequencies that are indeed indistinguishable within the detector
linewidth $|\omega_1-\omega_2|<\Gamma$. This accounts for the diagonal
feature on all plots. Apart from that, the harmonic oscillator
correlations lack any structure and are always above one,
corresponding to the expected bunching, whereas the two-level system
correlations mainly assume values below one, the expected
antibunching, furthermore in a nontrivial configuration of
detection~$\mathcal{C}$ with different frequencies that attempt to
maximise the overlap with the 1PS without entering the
indistinguishability bunching region:
\begin{equation}
  \label{antibunchingcondition}
  \mathcal{C}=\Big((|\omega_1-\omega_\sigma|<\Gamma_\sigma)\vee(|\omega_2-\omega_\sigma|<\Gamma_\sigma)\Big)\wedge(|\omega_1-\omega_2|>\Gamma)\,.
\end{equation}
The largest antibunching is thus obtained on the antidiagonal. In
practical terms, for a given detector linewidth, it is therefore
better to perform photon coincidences between the left and right
elbows of the 1PS, with a slight shift of both windows away from the
center, rather than between photons both coming from the central
peak. One can easily compute where exactly to place the filters by
evaluating $g_\Gamma^{(2)}[\sigma](\omega,-\omega)$ with
Eqs.~(\ref{eq:TueOct2143944CEST2012}) and~(\ref{twophotonLS}). For
instance, in the case $\Gamma=\Gamma_\sigma$, the minimum goes down
to~$2(\sqrt{2}-1)/5\approx 0.16$, for
$\omega_1=-\omega_2=\sqrt{(15\sqrt{2}-4)/31}\approx 0.75\,
\Gamma_\sigma$. This is the closest that the frequencies can get to
the central one without suffering from the indistinguishability
bunching. It is a considerable improvement, at the small cost of a
reduced signal, on measuring the correlations from the central peak
with a detector of the same linewidth, that provides an antibunching
of 0.5 only. When both detection windows are far from the peak, the
emitter looses its quantum character and behaves like its classical
counterpart, exhibiting bunching of its photons on the diagonal and
uncorrelated emission otherwise.

We complete the study of the correlations of a single two-level system
emitter by characterising the effect of inhomogeneous broadening. This
is relevant, for instance, in semiconductor systems (quantum dots)
where the solid state environment induces fluctuations and pure
dephasing on the levels. In this context, filtered two-photon
correlations have already shed light on the timescales and origin of
fluctuations in quantum dots~\cite{sallen10a}. Theoretically,
inhomogeneous broadening is introduced in the dynamics through another
Lindblad term of the form
$\frac{\gamma_\phi}{2}\mathcal{L}_{\ud{\sigma}\sigma}(\rho)$~\cite{laucht09b,gonzaleztudela10b}.
Pure dephasing occurs at the rate $\gamma_\phi$, diminishing coherence
in the system without affecting directly the populations. Its effect
on the 1PS is merely to increase the linewidth of the peak,
$\Gamma_\sigma=\gamma_\sigma + P_\sigma +\gamma_\phi$. For this
reason, it is typically difficult to measure a radiative lifetime from
spectroscopy and time-resolved measurements are usually invoked for
that purpose. For a given 1PS, however, the 2PS changes
quantitatively, which allows radiative and non-radiative contributions
to be measured directly from spectroscopy measurements. The 2PS
spectra without (c), and with (d), pure dephasing are compared in
Fig.~\ref{fig:1}.  While the statistics of the two-level system as a
whole is not affected by $\gamma_\phi$, its 2PS structure is. Namely,
pure dephasing extends the condition of
antibunching~(\ref{antibunchingcondition}) to a wider range of
frequencies and enhances the anticorrelations. This counter-intuitive
result is a manifestation in 2PS of the profitable effect of dephasing
for some quantum correlations~\cite{auffeves09a,auffeves10a}.

% Finally, in the limit of a dominant dephasing, the 2PS tends to a
% constant value independent of the frequencies considered,
% $\lim_{\gamma_\phi\rightarrow \infty}
% g_{\Gamma,\sigma}^{(2)}(\omega_1,\omega_2)=2({\gamma_\sigma+P_\sigma})/({\gamma_\sigma+P_\sigma+\Gamma})$.
 
\section{Coupled single mode emitters}
\label{doublemode}

The bare emitters already display a rich structure of two-photon
correlations. We now show that the next simple system, that of two
emitters coupled linearly, also exhibits two-photon correlations of
unsuspected complexity. Two coupled single-mode emitters account for,
or are at the heart of, a large gamut of quantum optical systems. We
will consider three paradigmatic cases, subject of intense theoretical
and experimental studies: two coupled harmonic
oscillators~\cite{reithmaier04a,kavokin_book11a}, two coupled
two-level systems~\cite{majer05a,gallardo10a,laucht10a,mueller12a} and
the Jaynes-Cummings model
\cite{jaynes63a,birnbaum05a,hennessy07a,nomura10a,lang11a,koch11a}
where one harmonic oscillator is coupled to a two-level system. All of
them are described by the following Hamiltonian:
\begin{equation}
  H_C=\omega_{O_1} \ud{O}_1 O_1 +\omega_{O_2} \ud{O}_2 O_2 +g (\ud{O}_1 O_2 +\ud{O}_2 O_1)\,,
\end{equation}
where, as in the previous section, $O_{1,2}$ describes two modes that
can be either an harmonic oscillator or a two-level system.  The dynamics of the system is completed
with the inclusion of decay and pumping through the master equation:
\begin{equation}
  \label{eq:MonNov19183152CET2012}
  \partial_t\rho=i[\rho,H_C]+[\frac{\gamma_{O_1}}2\mathcal{L}_{O_1}+\frac{\gamma_{O_2}}2\mathcal{L}_{O_2}+\frac{P_{O_1}}{2}\mathcal{L}_{\ud{O}_1}+\frac{P_{O_2}}{2}\mathcal{L}_{\ud{O}_2}](\rho)\,.
\end{equation}
In the rest of the text, for the sake of brevity, we will limit to the
situation where the first mode $O_1$ is detected and the other mode
$O_2$ is pumped ($P_{O_1}=0$). This reduces the number of parameters
and remains close to the experimental situation in cavity-QED where
typically the emitter is incoherently pumped and the system is
observed through the cavity emission.

\begin{figure}[t] 
  \centering
  \includegraphics[width=0.9\linewidth]{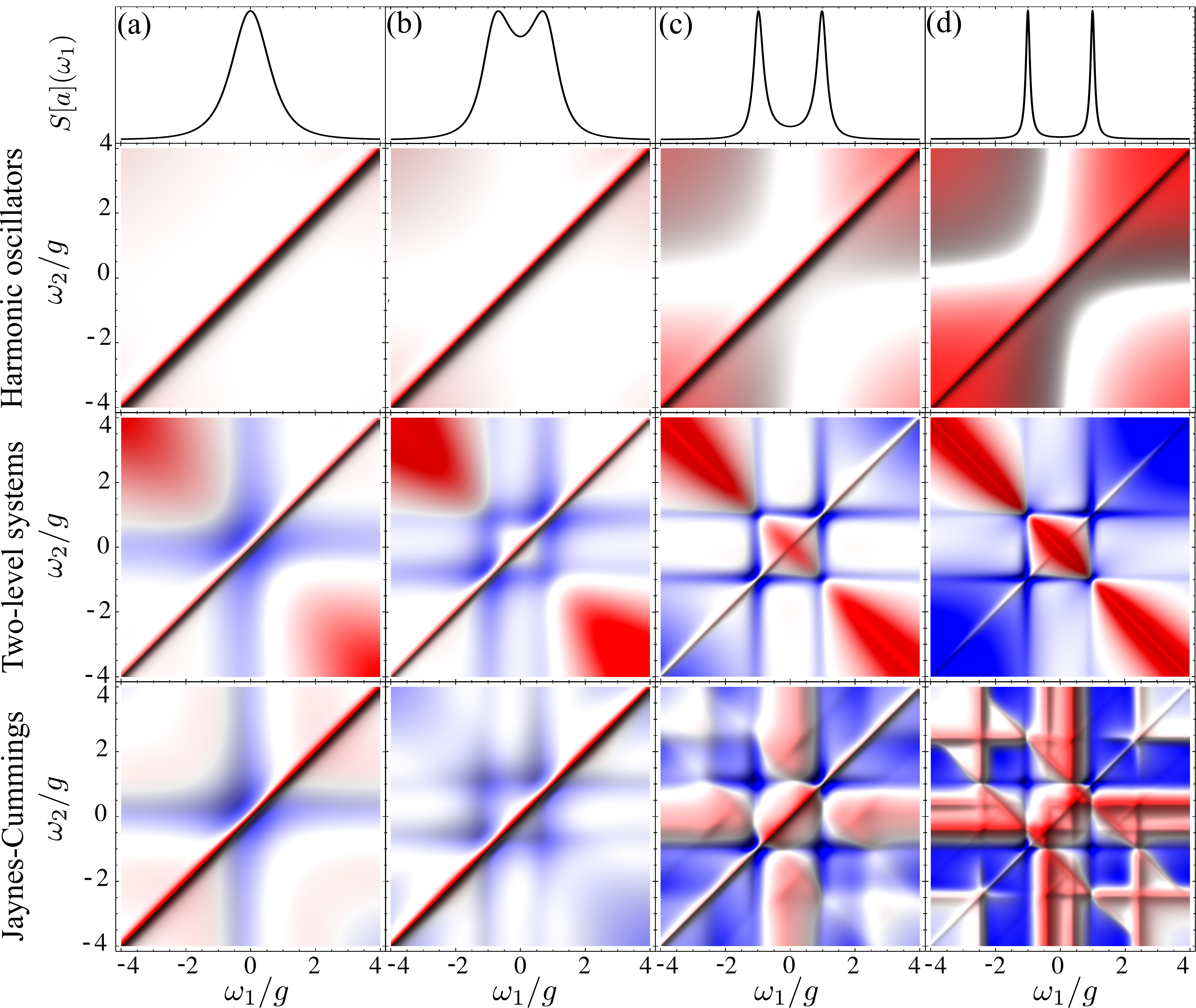}
   \caption{First row: 1PS that are, in the linear regime, identical
    for coupled harmonic oscillators, coupled two-level systems and
    the Jaynes--Cummings model. Second/third/fourth rows:
    corresponding 2PS, that are, on the contrary to the 1PS, markedly
    different. In the coupled two-level systems and the
    Jaynes--Cummings model, a logarithmic scale has been used to
    improve the visibility of all the features. Columns (a)-(d)
    correspond to increasing the quality of the coupling as
    $\gamma_{O_1}/g=4,\,2,\,1,\,0.5,\,0.1$. In all cases,
    $\Gamma=0.1g$, $P_{O_1}=0$ and $\gamma_{O_2}=0.001g$.}
  \label{fig:2}
\end{figure}

In Fig.~\ref{fig:2}, we show the 2PS of the $O_1$-mode for different
values of $\gamma_{O_1}$, from $4g$ (weak coupling) to $0.1g$ (strong
coupling), fixing the lifetime of the second mode to a small value
$\gamma_{O_2}=0.001g$, also on the basis of typical experimental
values. The detectors linewidths are set equal to that of the emitting
mode, $\Gamma=\gamma_{O_1}$, as previously. In the linear regime, when
$P_{O_2}\rightarrow 0$, all the models converge to the same 1PS (first
row), whereas their 2PS exhibit very different structures depending on
the nature of the emitters. When the losses overcome the coupling,
i.e., in the weak-coupling regime, the 1PS converges to that of the
bare mode~$O_1$. In the case of the coupled harmonic oscillators and
the coupled two-level systems, the 2PS has the form of the single
mode, cf.~Fig.~\ref{fig:1}(a) and (b), respectively. However, in the
case of the Jaynes-Cummings model, although the cavity mode is
detected, a statistics reminiscent of the two-level system is
observed, even in weak-coupling. This hints at the value of such
systems for quantum applications, such as a single-photon sources: the
(classical) cavity enhances the emission and collects the light while
still retaining the valuable quantum features of the (quantum)
emitter. Note that while juxtaposing the two filters on both sides of
the central peak was only an improvement in the case of the bare
two-level system, it is here mandatory, as the central peak is
strongly bunched.

As the quality of the coupling increases ($\gamma_{O_1}\rightarrow
0.1g$), the system enters in the strong-coupling with new dressed
states, or polaritons, emerging. We call $p_{\pm}$ the annihilation
operators for the upper~($+$) and lower polariton~($-$).  The 1PS,
still identical for the three coupling models, turns into the
so-called ``Rabi doublet'', with splitting
$2R=2\sqrt{g^2-[(\gamma_{O_1}-\gamma_{O_2})/4]^2}$ between the
polaritons (the splitting observed in PL is more complicated and is
given in Ref.~\cite{gonzaleztudela10a}). The 2PS, however, present,
again, different features that we discuss in turns.

\subsection{Coupled harmonic oscillators}

The coupled harmonic oscillators, that we denote $a$ and $b$, are
characterised by bunched correlations between any two frequencies, as
expected due to the bosonic character of both modes. Similarly to the
case of a single harmonic oscillator, the color-blind statistics
always remains thermal, for both the bare and dressed modes:
$g^{(2)}[a]=g^{(2)}[b]=g^{(2)}[p_+]=g^{(2)}[p_-]=2$. The cross
correlations, however, depend on the system parameters:
\begin{equation}
  \label{eq:FriOct5154620CEST2012}
  g^{(2)}[a;b]\equiv\frac{\mean{\ud{a}a\ud{b}b}}{\mean{n_a}\mean{n_b}}=2-\frac{P_a/\mean{n_a}+P_b/\mean{n_b}}{\Gamma_a+\Gamma_b}\,,
\end{equation}
(at resonance) and is always between 1 and 2. In the strong-coupling
regime, $g^{(2)}[a;b]\rightarrow1$, i.e., the two modes become
uncorrelated. The correlations between polaritons, $g^{(2)}[p_+;p_-]$,
are also close to 1 as long as polaritons are well defined (well
separated spectrally), due to the formation of their independent
polaritonic dynamics.

Coming back to frequency-resolved correlations, the 2PS of the
mode~$a$ (2nd row in Fig.~\ref{fig:2}) presents, in strong-coupling,
two hyperbolic dips where $g_\Gamma^{(2)}(\omega_1,\omega_2)$ drops to
its minimum value of~1.  The hyperbolas, given by the equation
\begin{equation}
  \label{bunchingcondition}
  \omega_1 \omega_2=-g^2\,,
\end{equation}
corresponds to the detected frequencies $(\omega_1,\omega_2)$ hitting
the polariton branches which, as a function of detuning~$\Delta$, read
$\omega_\pm(\Delta)=\Delta/2\pm\sqrt{g^2+(\Delta/2)^2}$. When one
detects a photon at $\omega_1$, the classical system does not provide
information on whether this photon comes from the main peak itself
(where the dressed state is located and the mode emits predominantly)
or from any other point on its tail. This photon is detected as if
emitted by the polariton being at precisely this energy. It has thus
the fitting correlation of the coupled oscillators with another photon
detected at the energy of the other polariton with the same detuning
as the would-be polariton accounting for the first photon.  The 2PS
thus evidence the existence of the full polariton dispersion via a
change in the statistics, even though the modes are at resonance in
the 1PS. This shows again the richer discriminative power of
two-photon spectroscopy.  The 2PS anticrossing is wider than the 1PS
one given that the distance between vertices $(\pm g,\mp g)$ is
$2\sqrt{2}g$ instead of $2g$. The renormalisation by $\sqrt{2}$ is due
to the fact that the information is carried at the two-photon level
instead of one in the photoluminescence.

\begin{figure}[t]
  \centering
  \includegraphics[width=0.9\linewidth]{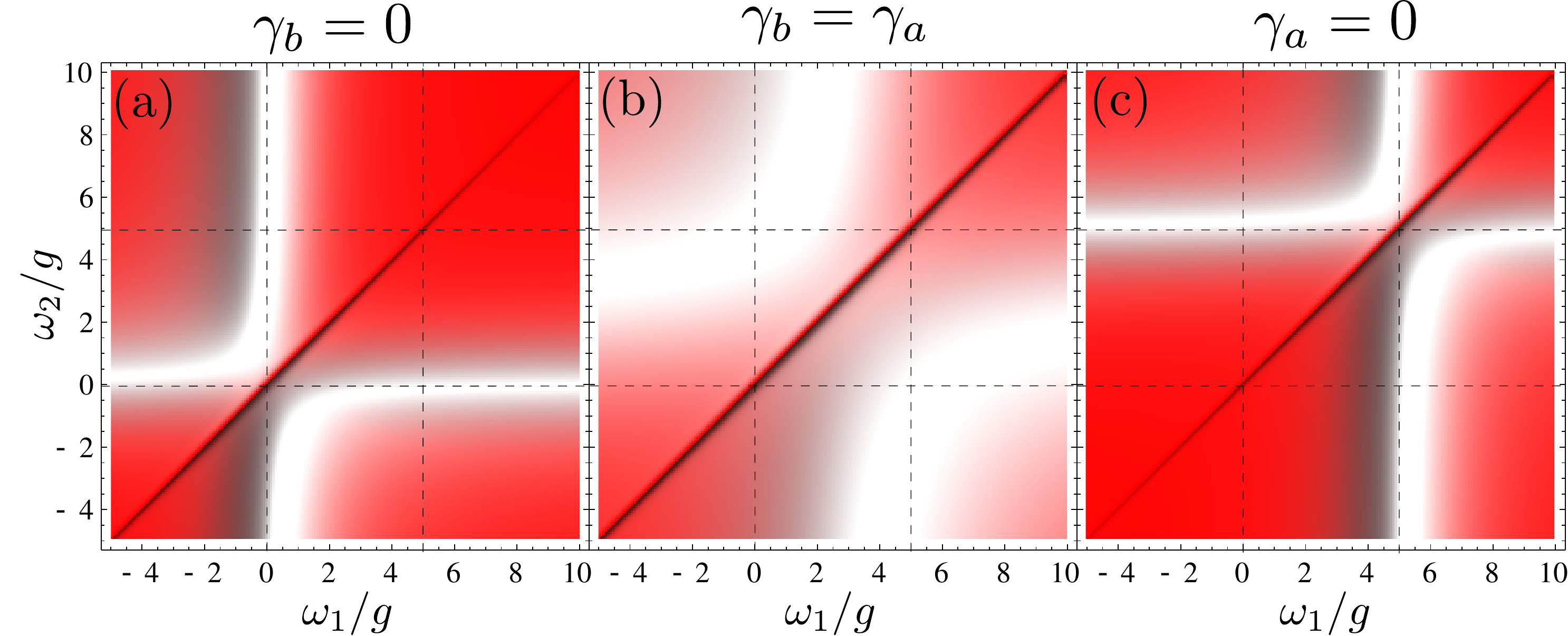}
  \caption{2PS for coupled and detuned harmonic oscillators. The bare
    states are at frequencies $\omega_a=0$ and $\omega_b=5g$ (dashed
    lines).  The total decay in the system is kept fixed to
    $\gamma_a+\gamma_b=0.1g$ while $\gamma_a$ and $\gamma_b$ vary as
    indicated above each plot. Other parameters: $P_a=0$, $P_b\ll
    \gamma_{a,b}$ and $\Gamma=0.1g$.}
  \label{fig:3}
\end{figure}

When the modes are detuned, the polariton correlation interferences
still form an hyperbola,
\begin{equation}
  \label{eq:TueSep4184740CEST2012}
  \Big (\omega_1-\omega_\mathrm{hyp}\Big) \Big(\omega_2-\omega_\mathrm{hyp}\Big)=-V_\mathrm{hyp}^2\,,
\end{equation}
but its center, $\omega_\mathrm{hyp}$, and vertex, $V_\mathrm{hyp}$,
depend on the set of relevant parameters of the system, namely
$\{P_a,P_b,\gamma_a,\gamma_b,\omega_a,\omega_b\}$, as shown in
Fig. \ref{fig:3}. In order to gain further insight into this
dependence, let us assume that polaritons are well defined and,
therefore, the 1PS is essentially the sum of two Lorentzian peaks
associated with each polariton, $S[a](\omega)\approx
L^a_{-}(\omega)+L^a_{+}(\omega)$. We can neglect the dispersive
contributions to the spectrum~\cite{laussy09a} thanks to the
negligible overlap between these peaks and assume complete
uncorrelation between the polaritons, $g^{(2)}[p_+;p_-] \approx 1$. In
a single particle picture, one can define a frequency dependent
operator for mode $a$, $\alpha(\omega)=\sqrt{L^a_{-}(\omega)}
p_{-}+\sqrt{L^a_{+}(\omega)} p_{+}$, using the relative spectral
weights,
\begin{equation}
  \label{eq:TueSep4184740CEST2012a}
  L^a_{\pm}(\omega)= \frac{l_{\pm}^a}{\pi}\frac{\gamma_{\pm}/2}{(\gamma_{\pm}/2)^2+(\omega-\omega_{\pm})^2}\,,
\end{equation}
that take into account the quantum state of the system through the
coefficients $l_\pm^a$~\cite{laussy09a}.  The parameters
$\gamma_{\pm}$ and $\omega_{\pm}$ are the full-width half-maximum and
positions of the polariton modes, $p_\pm$.  The 1PS is well
approximated by $\mean{\ud{\alpha}(\omega) \alpha (\omega)}$. In a
similar way,
$\mean{\ud{\alpha}(\omega_1)\ud{\alpha}(\omega_2)\alpha(\omega_2)\alpha(\omega_1)}$,
captures some of the features of the exact results in
Fig. \ref{fig:3}.  The correlations computed in this way are less well
reproduced than the 1PS as this approach neglects the multi-photon
dynamics. However, due to the linear nature of the system, they
produce interferences between the well defined polaritons with the
same hyperbola, Eq.~(\ref{eq:TueSep4184740CEST2012}), specified by:
\begin{eqnarray}
  \label{movement}
  \omega_\mathrm{hyp}\approx\frac{l_{-}^a}{l_{-}^a+l_{+}^a}\omega_{+}+ \frac{l_{+}^a}{l_{-}^a+l_{+}^a}\omega_{-}\, ,\\
  V_\mathrm{hyp}\approx \Re\sqrt{g^2-\Big(\frac{4l_{-}^a l_{+}^a}{(l_{-}^a+l_{+}^a)^2}\Big)^2\Big(\frac{\Gamma_a-\Gamma_b}{4}+i\frac{\omega_a-\omega_b}{2}\Big)^2}\, .
\end{eqnarray}

In Fig. \ref{fig:3}, we exemplify how the quantum state affects the
anticrossing of the 2PS for a detuned situation, where
$\omega_-\approx \omega_a=0$ and $\omega_+\approx \omega_b=5g$. For
simplicity, only one mode is pumped in the linear regime ($P_a=0$,
$P_b\ll \gamma_b$).  The total decay of the system is fixed
($\gamma_a+\gamma_b=0.1g$) whereas the relative value of the two rates
is varied to alter the quantum state of the system. In the asymmetric
case where $\gamma_b=0$, Fig. \ref{fig:3}(a), the normalized weight is
$l_{+}^a\approx 0$ and, in accordance with Eq.~(\ref{movement}), the
center of the anticrossing $\omega_\mathrm{hyp}\approx \omega_a$ with
a vertex $V_\mathrm{hyp}\approx g$. In the symmetric situation where
$\gamma_a=\gamma_b$, Fig. \ref{fig:3}(b), both weights are equal,
$l_{+}^a\approx l_{-}^a$, yielding a balanced position of the
anticrossing between the two modes, $\omega_\mathrm{hyp}\approx
\frac{\omega_a+\omega_b}{2}$, and a larger splitting given by the same
Rabi splitting as the 1PS, $V_\mathrm{hyp}\approx
\sqrt{g^2+(\omega_a-\omega_b)^2/4}$. Finally, in Fig. \ref{fig:3}(c)
where $\gamma_a=0$, the quantum state with $l_{-}^a\approx 0$, yields
$\omega_\mathrm{hyp}\approx \omega_b$ and the same vertex as in the
case of Fig. \ref{fig:3}(a), $V_\mathrm{hyp}\approx g$.

\subsection{Coupled two-level systems}

The coupled two-level systems (3rd row in Fig.~\ref{fig:2}) differs
from the coupled harmonic oscillators in much the same respect than
the two-level system does from the harmonic oscillator. There are now,
unsurprisingly, regions of antibunching and a richer pattern of
correlations.

As the system enters strong-coupling, a grid of two vertical and two
horizontal lines of antibunching appear, corresponding to one
frequency fixed at one of the peaks at $\pm R$ and the other frequency
scanning through the 1PS. The stronger antibunching is obtained at
their crossing. There is also the diagonal of indistinguishability
bunching. Finally, a neat feature revealed by the 2PS is the strong
superpoissonian antidiagonal, when $\omega_1+\omega_2=0$. This is the
first occurrence of an important and recurrent pattern in two-photon
spectroscopy: these correlations correspond to direct two-photon
relaxation from the doubly occupied state
$\ket{1,1}\rightrightarrows\ket{0,0}$, through a virtual
(off-resonant) intermediate state. As the energy of this auxiliary
step is not constrained, the feature appears on an entire line. Only
thanks to the coupling can one two-level system effectively emit two
excitations at the same time (within the detector time
window~$1/\Gamma$). We use the denomination of ``\emph{leapfrog}'' to
label this type of relaxation, as the intermediate rung is not
populated and the system effectively jumps over it by emitting two
photons simultaneously. It becomes more prominent as the detector
width, $\Gamma$, decreases and the time uncertainty of the measurement
increases. This is a fundamental process that is common to all quantum
nonlinear systems. It supports the idea that the emission of two
identical photons can be enhanced (and selected) by placing such
systems in a high-$Q$ cavity~\cite{delvalle10a,delvalle11d,ota11a}. In
the simple system that is the two-coupled two-level systems, it
appears in isolation. We discuss it in more extent in the next
paragraph where it appears along with similar processes, making
sharper its own features.

\begin{figure}[t] 
  \centering
  \includegraphics[width=0.8\linewidth]{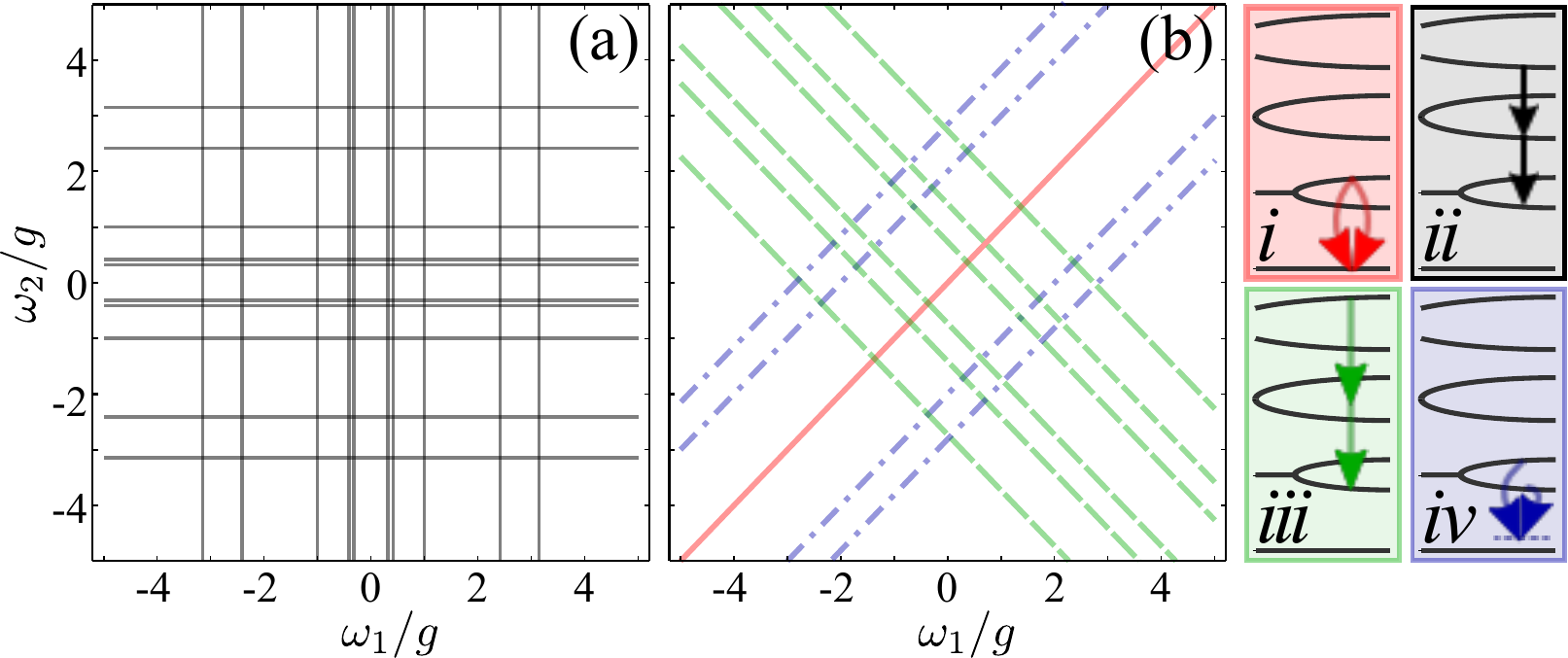}
  \caption{Patterns of two-photon correlations in the Jaynes--Cummings
    model up to the third rung of excitation (low pumping regime). (a)
    The grid of horizontal and vertical lines (solid black)
    corresponds to correlations between real states, when one
    frequency is pinned at a transition in the ladder that is
    correlated with emission anywhere else in the system resulting in
    resonances when matching with a cascade (bunching) or an
    incompatible process like emission from the other type of
    polariton (antibunching). (b) The grid of antidiagonal lines
    (dashed green) corresponds to leapfrog processes where two-photon
    transitions take place from one rung of the ladder to another one
    two steps below, jumping over the intermediate rung. The grid of
    diagonal lines (except the central one, dashed-dot blue)
    correspond to polariton-to-virtual-state antibunching.  The
    ``indistinguishability bunching'' (diagonal red line) is common to
    all 2PS.  These two panels overlapping emerge clearly in the best
    systems, cf.~Fig.~\ref{fig:2}(c)--(e), up to the second rung, and
    Fig.~\ref{fig:5}(c),(d).  Panels $i$--$iv$ on the right are
    sketches of the underlying processes responsible for these
    correlations (with the same color code).}
  \label{fig:4}
\end{figure}

\subsection{Jaynes-Cummings model}

In mixing the types of emitters, one harmonic oscillator (cavity) and
one two-level system (atom, quantum dots, superconducting qubit\dots),
a much richer dynamics is unravelled by the 2PS (4th row in
Fig.~\ref{fig:2}). In strong-coupling, a ladder-type level structure
(shown in Fig.~\ref{fig:4}) arises thanks to the nonlinear splitting
$2\sqrt{(\sqrt{n}g)^2-[(\gamma_a-\gamma_\sigma)/4]^2}$ between the
states with $n$ excitations~\cite{delvalle09a}. The one-photon
relaxations between these states give rise to a rich multiplet
spectrum~\cite{laussy09b}, and to a considerably richer set of
two-photon relaxations, that we now discuss in details.

The simplest and most straightforward type of two-photon correlations
is the one that precisely follows from sequences of two one-photon
relaxations, with each photon emitted from one state of the
Jaynes-Cummings ladder to another. Depending on the type of
relaxation, a given type of correlation is expected. For instance, at
low pumping, detecting emission from the two types of polaritons in
any given rung of the ladder leads to antibunching, as the
de-excitation went one way or the other. Similarly, detecting emission
in cascades from the ladder in consecutive events (as depicted in
Fig.~\ref{fig:4}$ii$) leads to bunching. Such processes are the
extensions to all possible rungs of the ladder of the correlations
observed with the coupled two-level systems. In fact, when the system
is not in very strong-coupling, and thus makes it difficult to resolve
the dynamics of the higher rungs, the 2PS of the Jaynes-Cummings
resembles that of the coupled two-level systems, with four
antibunching dips appearing at $\omega_1$ or $\omega_2=\pm R$,
corresponding to the situation depicted in Fig.~\ref{fig:4}$i$.

\begin{figure}[t]
  \centering
  \includegraphics[width=0.75\linewidth]{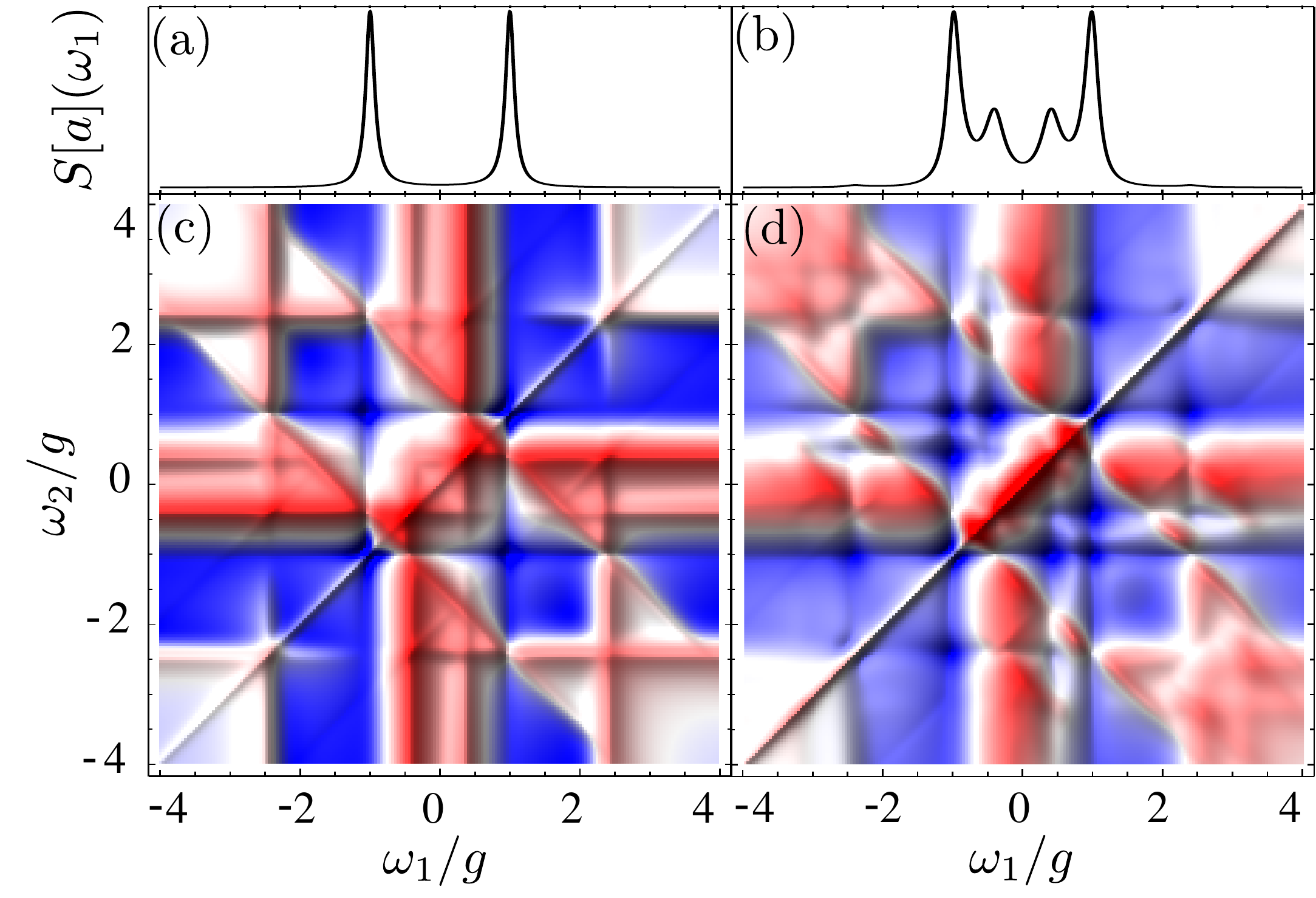}
  \caption{1PS (first row) and 2PS (second row) for Jaynes-Cummings, going from
    vanishing pumping in (a), (c), to the nonlinear regime in (b),
    (d), $P_\sigma=0.05g$. Second rung transitions appear in the 1PS
    inside the Rabi doublet and new features involving the third rung
    of excitation appear in the 2PS.}
  \label{fig:5}
\end{figure}

As the system gets deeper into strong-coupling, more such correlations
from transitions between different rungs are revealed. They are given
by $\omega_1=E_n\pm E_{n-1}$ and $\omega_2$ free (vertical lines) and
their counterpart $\omega_1\leftrightarrow\omega_2$ (horizontal
lines).  All together they constitute the grid of horizontal and
vertical lines shown in black in Fig.~\ref{fig:4}(a). This pattern is
neatly reproduced by the actual 2PS (see Fig.~\ref{fig:5}, for the
best system) with enhancement/suppression of the correlations at every
crossing point depending on whether the combination of frequencies
corresponds to consecutive/non-consecutive
processes~\cite{delvalle12a}.

There is also the now familiar diagonal of indistinguishability
bunching, in red in Fig.~\ref{fig:4}(b), and other diagonal and
antidiagonal features, better resolved in the best systems. These
correspond to emission through virtual states during the short
interaction time with the detector, $\sim 1/\Gamma$, and have, to the
best of our knowledge, not been reported up to now. The two
antidiagonal lines at $\omega_1+\omega_2=2\omega_a\pm E_2\approx
\pm\sqrt{2}g$ (dashed green lines in Fig.~\ref{fig:4}(b)) are the
leapfrog processes already encountered with the coupled two-level
systems: a two-photon transition from the second rung directly to the
ground state, transiting through a virtual intermediate state which
unspecified energy allows this process to be detected at any point on
a line. This process is depicted in Fig.~\ref{fig:4}$iii$. The
correlation for such a direct two-photon emission is, naturally,
bunched.

Increasing the excitation power, the higher rungs get populated and
the additional leapfrog process
$\ket{3\pm}\rightrightarrows\ket{1\pm}$ can be observed. The four
satellite antidiagonal lines that result, satisfying
$\omega_1+\omega_2= 2\omega_a \pm (E_3 - E_1)$ or $\pm (E_3 + E_1)$,
are plotted in Fig.~\ref{fig:4}(b) and materialised in the 2PS in
Fig.~\ref{fig:5}(d). These lines are more visible in some regions (of
strong bunching) but upon closer inspection, a deviation of the
statistics due to such processes is realized everywhere.

A feature not observed in the previous cases is the secondary diagonal
lines at $\omega_1-\omega_2= \pm 2R$, that meets with the antibunching
dips at the first rung resonances. They are plotted as dotted-dashed
blue lines in Fig.~\ref{fig:4}(b). In the actual 2PS, these are globally
weaker and more clearly distinguished in some regions only, like the
leapfrog lines from the higher rungs, and are also realized everywhere
if looked at closely.  Unlike leapfrog processes or
indistinguishability bunching, however, they do exhibit an
antibunching tendency as compared to the surrounding region (even if
the overall correlation remains bunched).  They correspond to the
anticorrelated emission from both polaritons of the first rung into
the same virtual final state, as depicted in
Fig.~\ref{fig:4}$iv$. These lines become more pronounced when the mode
under observation is also the one that is excited (as stated
previously, we consider the situation where one mode is excited and
the other one is observed). In such a pumping configuration, strong
polariton-to-virtual-state correlations also appear in the coupled
two-level systems (not shown) while, as seen in Fig.~\ref{fig:2},
third row, they are absent in the other pumping-detection scheme. This
suggests that the interference between relaxations from $\ket{+}$ and
$\ket{-}$ plays a role.  They are thus not peculiar to the
Jaynes-Cummings model but to any strongly coupled, nonlinear
system. Higher excitation, makes higher order diagonal lines appear,
corresponding to higher rung dressed states, such as
$\omega_1-\omega_2=\pm 2 E_n$. Such diagonals from the second rungs
are also visible in Fig.~\ref{fig:5}(d). In this figure, the effect of
increasing pumping is shown to ``melt'' the pattern of correlations as
more real states are populated and more features distributed over the
2PS.

All these results show that two-photon spectroscopy unfolds the
mechanisms of relaxation that are not apparent in
photoluminescence. This can be used to characterize a system or
exploit quantum correlations.

\section{Lasing and the Mollow triplet}
\label{sec:ThuJun14164126CEST2012}

\begin{figure}[t] 
  \centering
  \includegraphics[width=\linewidth]{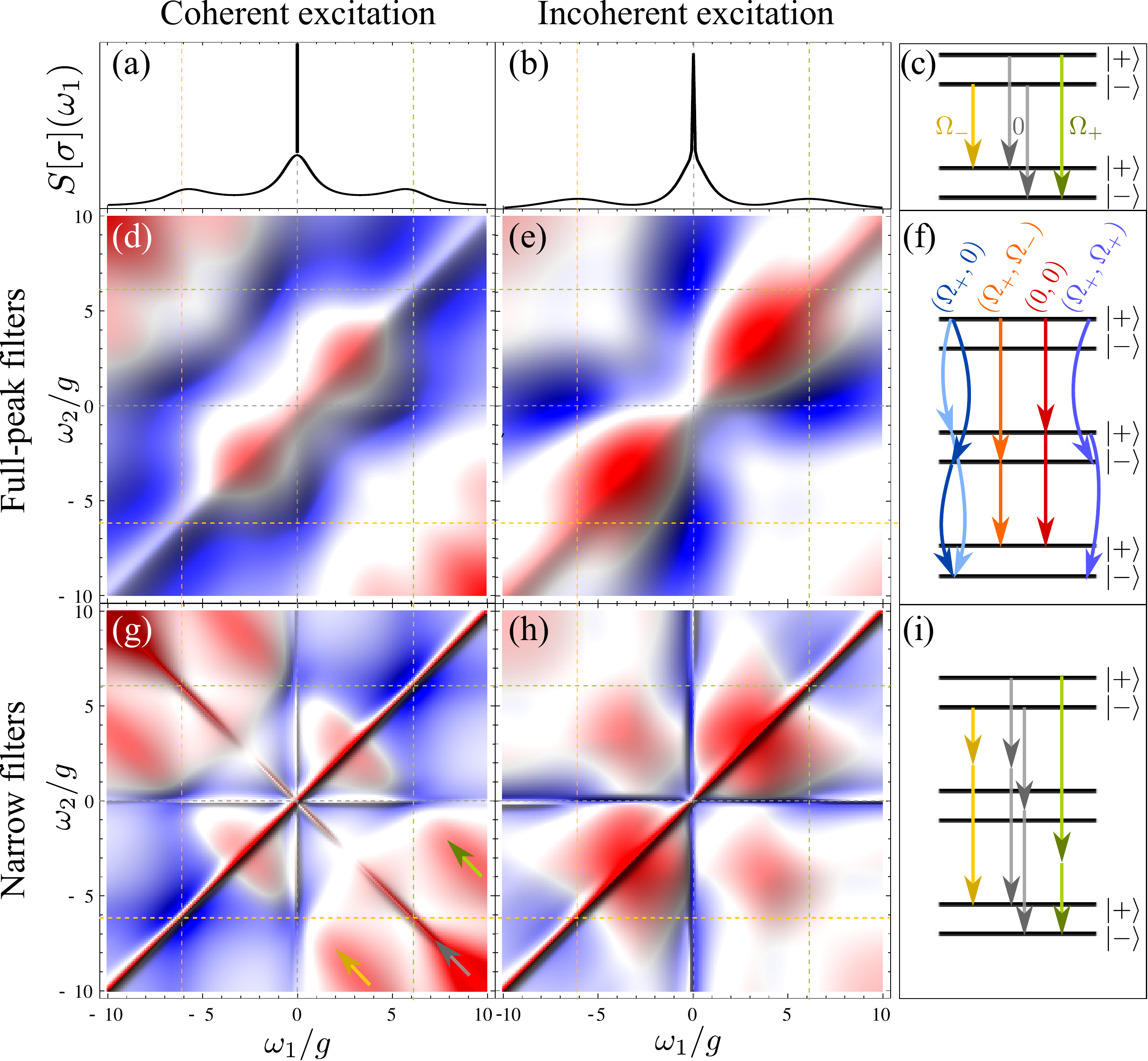}
   \caption{Normalised 1PS (first row) and 2PS (second and third rows)
    for the Mollow triplet under coherent (left) and incoherent
    excitation (right). Parameters are chosen so that the sidebands
    are at the same positions $\Omega_\pm=\pm 6.16 g$ and have the
    same broadening with $P_\sigma=2 g$ and
    $\mean{n_a}=9.5$. Vertical/horizontal gridlines mark the positions
    of the upper (green), lower (orange) and central black (peaks),
    arising from the transitions between dressed states depicted in
    (c). In (d) and (e) the detector width, $\Gamma=3 P_\sigma/2$, is
    wide enough to filter full spectral peaks. Clear correlations
    appear for the pairs of transitions in (f). In (g) and (h) the
    detector width, $\Gamma=P_\sigma/2$, is smaller than the spectral
    peaks. In this case, the dominant feature is the \emph{leapfrog}
    triplet of antidiagonal lines, corresponding the to two-photon
    de-excitation processes sketched in (i).}
  \label{fig:6}
\end{figure}
 
We have just discussed the effect of increasing the excitation power
in the case of the Jaynes-Cummings model: higher rungs get populated
and give rise to additional relaxation processes, either from real
states that involve photon by photon de-excitations or through
two-photon emission that involve intermediate virtual states. Further
increasing $P_\sigma$ may bring the Jaynes-Cummings system into the
lasing regime~\cite{mu92a,delvalle11a} where the optical field becomes
coherent and $g^{(2)}[a]\rightarrow 1$. The cavity 1PS reduces to a
single line that narrows with increasing occupancy of the mode as
$\gamma_\mathrm{L} \approx g^2 /(2\gamma_a
\mean{n_a}^2)$~\cite{poddubny10a}, where $\mean{n_a}\approx P_\sigma
/(2\gamma_a)$. The inverse quantity, $1/\gamma_\mathrm{L}$,
corresponds to the coherence time of the laser. More interesting is
the 1PS of the two-level system (or quantum emitter), which converges
to a {\it Mollow triplet}~\cite{mollow69a}, with some specificities of
its own in the cavity QED version as compared to resonance
fluorescence due to the incoherent nature of the
excitation~\cite{delvalle10d}. Therefore, in this discussion we
analyse the emission directly from the two-level system; this is the
mode to which the sensors are now connected.

A Mollow triplet arises when a two-level system is driven by intense
laser light. The splitting from very high rungs of the Jaynes-Cummings
ladder becomes homogeneous giving rise to similar dressed states,
$\ket{+}=c\ket{0}-s\ket{1}$ and $\ket{-}=s\ket{0}+c\ket{1}$, with
$\ket{0}$ the ground state and $\ket{1}$ the excited state of the
two-level system.  The ladder operator can be decomposed into the four
possible transitions,
$\sigma=\ket{0}\bra{1}=c^2\ket{+}\bra{-}-cs\ket{+}\bra{+}+cs\ket{-}\bra{-}-s^2\ket{-}\bra{+}$. The
first and last terms give rise to sidebands in the 1PS at positions
$\Omega_\pm \approx \pm 2g\sqrt{\mean{n_a}}$ (at resonance), as showed
schematically in Fig.~\ref{fig:6}(c).

A standard way to describe the laser excitation is through the
semiclassical Hamiltonian term $H_\mathrm{L}=\Omega_\mathrm{L} (\sigma
+\ud{\sigma})$ that couples the classical term $\Omega_\mathrm{L}$
(describing a macroscopic laser field) and the quantum term $(\sigma
+\ud{\sigma})$ (describing the quantum emitter)~\cite{mollow69a}. In
this approximation, the laser has an infinite coherence time
($\gamma_\mathrm{L}=0$) with intensity
$\mean{n_a}=\Omega_\mathrm{L}^2/g^2$ and $\Omega_\pm=\pm
2\Omega_\mathrm{L}$. This is a successful description of the
experimental observations, that report the theoretical lineshape,
shown in Fig.~\ref{fig:6}(a). A fully quantised picture for the
one-emitter laser, such as the Jaynes-Cummings model under incoherent
pumping, provides the counterpart lineshape in
Fig.~\ref{fig:6}(b). Parameters were chosen so that $\mean{n_a}$
coincide in both cases (to $9.5$) and the three peaks match in
position and broadening, with $P_\sigma=2 g$ for the
Jaynes-Cummings. They are not expected to match exactly since in the
former case the coherence of the laser is perfect and independent of
the emitter, while in the latter case, the coherence of the
light-field is established by the two-level emitter itself. They do
share, however, essentially the same phenomenology, with the
fundamental difference of quantisation or not of the light-field
between the two. In the lineshapes, this causes the important
difference of a finite linewidth $\gamma_\mathrm{L}$ for the Rayleigh
scattering peak on top of the two-level system triplet in the
Jaynes-Cummings model, while it is a Dirac $\delta$ function in the
conventional Mollow triplet theory due to the infinite coherence time
of the laser. We will see that, although this is a rather innocent
departure in the 1PS, it has dramatic consequences for the photon
statistics.

\begin{figure}[t] 
  \centering
  \includegraphics[width=\linewidth]{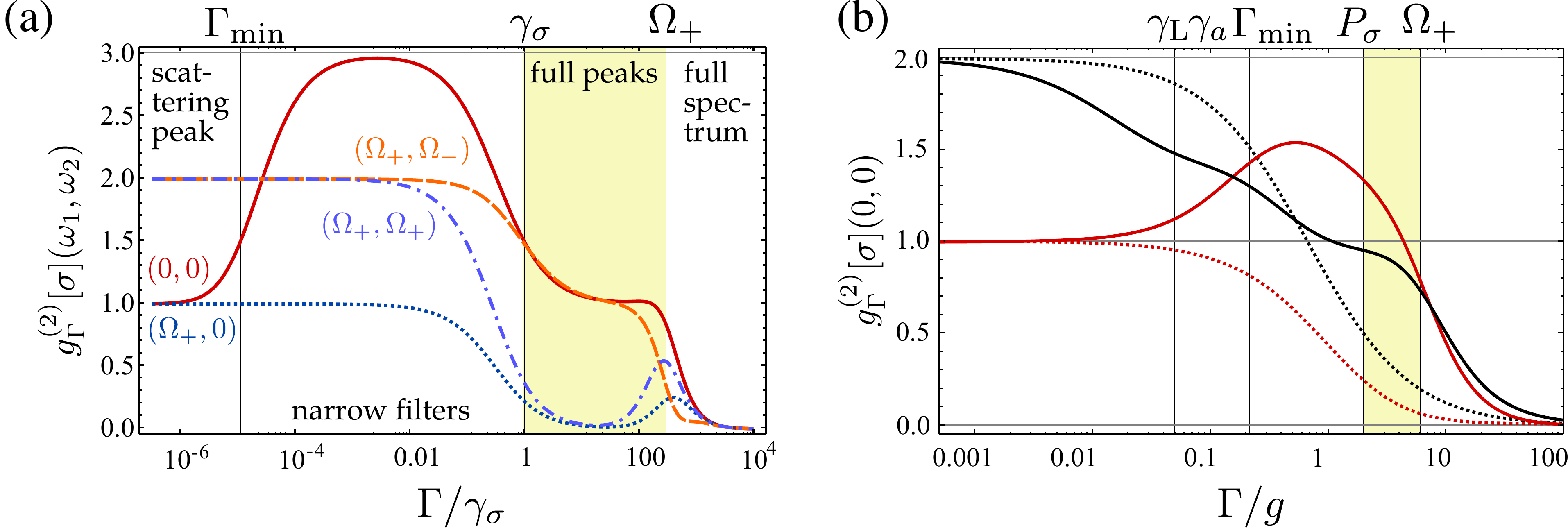}
  \caption{Two-photon correlators for frequencies of interest of the
    Mollow triplets as a function of the detector linewidth,
    $\Gamma$. (a) Correlations of a two-level system driven by a large
    coherent pump ($\Omega_\mathrm{L}=150\gamma_\sigma$) for the four
    most relevant $(\omega_1,\omega_2)$ configurations: $(0,0)$ in a
    solid red line, $(\Omega_+,\Omega_-)$ in dashed orange,
    $(\Omega_+,\Omega_+)$ in dotted-dashed blue and $(\Omega_+,0)$ in
    dotted blue. Four regions of detection can be defined (separated
    by vertical gridlines).  Correlations from the central peak (solid
    red) exhibit a bunching region of
    $g^{(2)}_\Gamma[\sigma](0,0)\approx 3$ at small $\Gamma$ when the
    coherent scattering is a significant fraction of the filtered
    light. Vanishing $\Gamma$ yields the statistics of the classical
    laser since the semiclassical zero linewidth cannot display
    indinstinguishability bunching. (b) Comparison of the central peak
    correlations, $(0,0)$ (solid red), in the case of an incoherently
    pumped two-level system in free space (dotted black) and in a
    cavity (solid black). The laser intensity here is lower:
    $\Omega_\pm=\pm 6.16g$ as in Fig.~\ref{fig:6}. The linear regime,
    $\Omega_\mathrm{L}\ll \gamma_\sigma$, is plotted with a dotted red
    line as a reference in units of $\gamma_\sigma=g$. When the photon
    field is quantized, the correct limit
    $g^{(2)}_\Gamma[\sigma](0,0)=2$ when $\Gamma\rightarrow0$ is
    recovered.}
  \label{fig:7}
\end{figure}

We now describe the 2PS of these two cases. Correlations between full,
well separated peaks of the semiclassical Mollow triplet have been the
subject of intense experimental~\cite{aspect80a,schrama91a,ulhaq12a}
and
theoretical~\cite{cohentannoudji79a,dalibard83a,reynaud83a,knoll84a,arnoldus84a,knoll86a,nienhuis93a,joosten00a,bel09a}
efforts (note that the Mollow triplet of a cavity QED system in
strong-coupling under incoherent pumping has still not be reported
experimentally at the time of writing). Recently, the full 2PS (more
specifically a closely related generalised Mandel $Q$ parameter) of
resonance fluorescence was obtained~\cite{bel09a} within the
generating function formalism, via single-molecule photon counting
statistics with spectral resolution. The 2PS for both cases are
compared in Fig.~\ref{fig:6}. They are smoother than in the quantum
regime, Fig.~\ref{fig:2}, and with patterns more dominantly aligned
along diagonals and antidiagonals rather than horizontal and vertical
lines, implying that the system is close to a macroscopic, classical
state as opposed to being dominated by correlations between well
defined real states.  One can understand the main features of these
plots again through a decomposition into two-photon dressed state
transitions:
\begin{eqnarray}
  \label{eq:FriSep14164821CEST2012}
  \sigma\sigma&=s^3c\ket{-}\Big(\bra{+}+\rangle-\bra{-}-\rangle\Big)\bra{+}+c^2s^2\ket{+}\Big(\bra{+}+\rangle-\bra{-}-\rangle\Big)\bra{+}\nonumber\\
  -&c^3s\ket{+}\Big(\bra{+}+\rangle-\bra{-}-\rangle\Big)\bra{-}-c^2s^2\ket{-}\Big(\bra{+}+\rangle-\bra{-}-\rangle\Big)\bra{-}\,.
\end{eqnarray}
Each of the $2^3$ terms represents a possible deexcitation path and
all together, they interfere destructively to give rise to the
expected total antibunching $\sigma \sigma=0$. The first line
corresponds to the four cascade transitions depicted in
Fig.~\ref{fig:6}(f), that start from $\ket{+}$:
\begin{itemize}
\item $(\Omega_+,0)$: One would expect this combination to produce
  bunched or cascaded photons but, as it corresponds to two different
  paths with amplitudes of probability with opposite signs (first two
  terms in Eq.~(\ref{eq:FriSep14164821CEST2012})), destructive
  interference leads to $g_\Gamma^{(2)}[\sigma](\Omega_+,0)=0$. Such
  \emph{debunching} effect occurs within the detector timescale
  $1/\Gamma$~\cite{schrama91a,nienhuis93b}.
\item $(\Omega_+,\Omega_-)$: Their cascade configuration produces a
  strong bunching, with a well defined time order that depends on the
  detuning with the laser~\cite{dalibard83a}.  However, at resonance,
  destructive interference debunches again the statistics to
  $g_\Gamma^{(2)}[\sigma] (\Omega_+,\Omega_-)=1$.
\item $(0,0)$: A similar situation of bunching and interference leads
  to $g_\Gamma^{(2)}[\sigma](0,0)=1$.
\item $(\Omega_+,\Omega_+)$: photons from the same sideband do not
  form a cascade so their simultaneous emission is suppressed (despite
  the indistinguishability bunching):
  $g_\Gamma^{(2)}[\sigma](\Omega_+,\Omega_+)=0$.
\end{itemize}
These ideal correlations are shown in Fig.~\ref{fig:7}(a) in the full
peak detection region (in yellow), where $\Omega_\mathrm{L}\gg
\Gamma_\sigma$.  With a less intense laser such as than in
Fig.~\ref{fig:6}, the features are still visible but slightly tempered
due to the overlap between peaks, as is the case in
experiments~\cite{ulhaq12a}.

\begin{figure}[t] 
  \centering
  \includegraphics[width=0.991\linewidth]{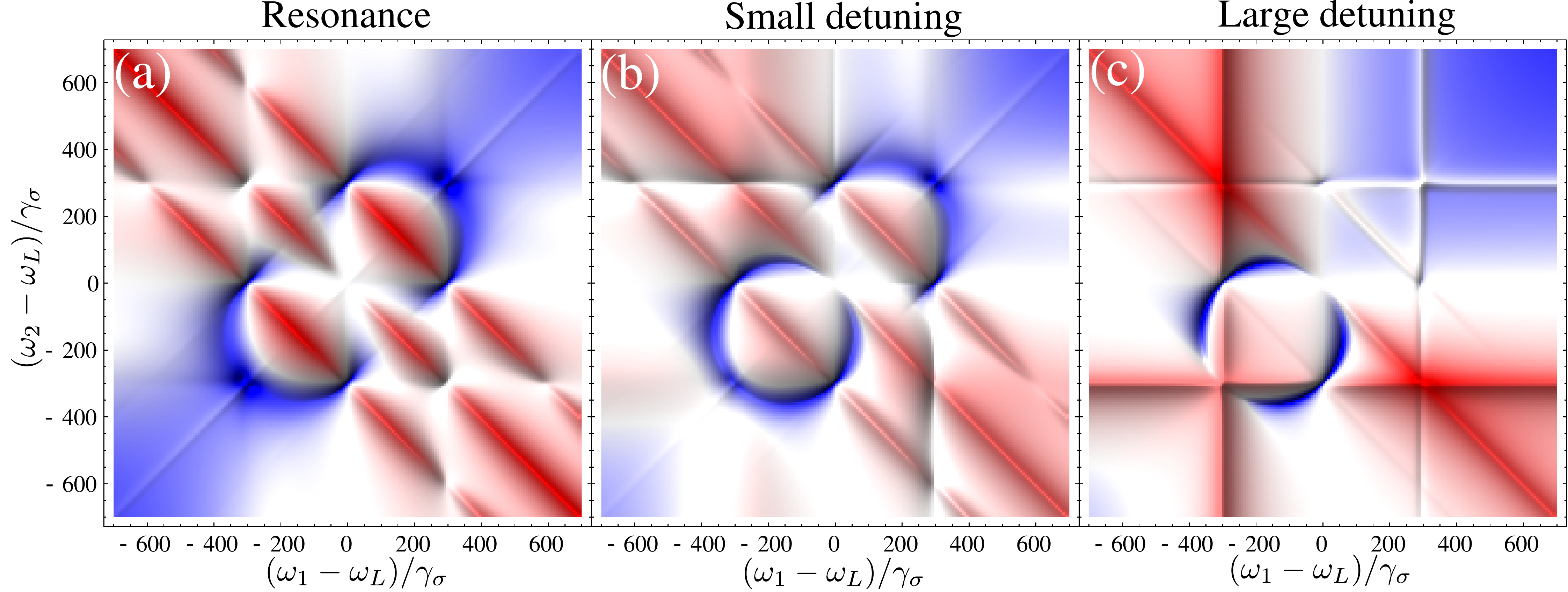}
  \caption{2PS for a two-level system under coherent excitation deep
    in the Mollow triplet regime
    ($\Omega_+=\sqrt{(2\Omega_\mathrm{L})^2+(\omega_\sigma-\omega_\mathrm{L})^2}=300\gamma_\sigma$)
    in the cases where (a) the laser is on resonance with the emitter
    ($\Omega_\mathrm{L}=150\gamma_\sigma$, as in Fig.~\ref{fig:7}(a))
    and (b) the laser is slightly detuned
    ($\Omega_\mathrm{L}=100\gamma_\sigma$) or (c) largely detuned
    ($\Omega_\mathrm{L}=10\gamma_\sigma$). In contrast with the 1PS,
    the 2PS is asymmetric with detuning. Also, in this regime where a
    highly coherent field is involved, a new type of pattern emerges
    in the correlations, namely, circles.}
  \label{fig:8}
\end{figure}

Thanks to the ease of use of our general solution, we are able to
compare the semiclassical model for the laser~\ref{fig:6}(d) with the
Jaynes-Cummings model in the lasing regime~\ref{fig:6}(e). In these
qualitatively similar plots, the interference effect described above
is even more evident as it extends beyond the four points
$(\Omega_\pm,0)$, forming two blue rings of antibunched
correlations. These interference rings are more clear in the strongly
driven situation depicted in Fig.~\ref{fig:8}. Their origin, as for
the leapfrogs, stems from the uncertainty introduced by including the
detector physics, which extends the interference appearing at the
points $(\Omega_\pm,0)$ and $(\Omega_\pm,\Omega_\pm)$ to two
circles. At the center of these interference rings,
($\Omega_\pm/2,\Omega_\pm/2$), two red spots of enhanced emission
appear from the concentration of leapfrog outer lines in the
Jaynes-Cummings model, Eqs.~(\ref{eq:WedAug8161036CEST2012}). With a
narrower detector, leapfrogs are enhanced due to the longer
uncertainty in the time of two-photon emission, as shown in
Fig.~\ref{fig:6}(g) and (h). In the same way that the Jaynes-Cummings
multiplets from high rungs of excitation converge to the Mollow
triplet in the 1PS, the antidiagonal leapfrog lines in the 2PS
converge to a \emph{leapfrog triplet} at:
\begin{equation}
  \label{eq:WedAug8181318CEST2012}
  \omega_1+\omega_2 \approx 0\,, \quad \omega_1+\omega_2 \approx \Omega_\pm\,,
\end{equation}
as schematically depicted in Fig.~\ref{fig:6}(i). They are more
pronounced in Fig.~\ref{fig:6}(g), as the perfect laser approximation,
$H_L$, generates dressed states homogeneously split for all
intensities $\Omega_\mathrm{L}$. 

If the detectors are narrower than the peaks, $\Gamma<\Gamma_\sigma$,
the interference effect described above disappears (see
Fig.~\ref{fig:7}).  Furthermore, as has been previously discussed,
when $\Gamma$ is smaller than any peak width, photons are detected
from any point of the dynamics and appear as uncorrelated. For
different frequencies this means $\lim_{\Gamma \rightarrow
  0}g_\Gamma^{(2)}[\sigma](\omega_1,\omega_2)=1$ while for equal
frequencies, $\lim_{\Gamma \rightarrow
  0}g_\Gamma^{(2)}[\sigma](\omega,\omega)=2$ due to the
indistinguishability of the photons. In this limit, the statistics
observed can no longer be attributed to the system, but to the
detection process. The two-level system excited incoherently (with or
without the cavity) always recovers this limit correctly as, in the
Jaynes-Cummings model, the smallest width, $\gamma_\mathrm{L}$ is
still finite and sets a clear lower boundary (see
Fig.~\ref{fig:7}(b)). In contrast, we have a totally different result
under coherent excitation of a perfect laser whenever
$\omega_1+\omega_2=2\omega_\mathrm{L}=0$: $\lim_{\Gamma\rightarrow 0}
g_\Gamma^{(2)}[\sigma](0,0)=1$ and $\lim_{\Gamma\rightarrow 0}
g_\Gamma^{(2)}[\sigma](\Omega_+,\Omega_-)=2$. Due to the implicit
assumption of zero-linewidth for the laser field, it is not possible
in these cases for the detector linewidth to be thinner and reach the
uncorrelated limit. Instead, the limit $\Gamma\rightarrow 0$ isolates
completely the Rayleigh peak so correlations become exactly 1, i.e.,
those of the laser. This occurs with a detector linewidth below
$\Gamma_\mathrm{min}$, defined as that at which the coherent part of
the filtered 1PS is as large as the incoherent part,
\begin{equation}
  \label{eq:MonSep17200333CEST2012}
  S^\mathrm{coh}_{\Gamma_\mathrm{min} }[\sigma](0)=S^\mathrm{incoh}_{\Gamma_\mathrm{min} }[\sigma] (0)\,,
\end{equation}
where $S_\Gamma [\sigma] (\omega)=S^\mathrm{coh}_\Gamma [\sigma]
(\omega)+S^\mathrm{incoh}_\Gamma [\sigma] (\omega)$. In the limit of
intense lasing, $\Gamma_\mathrm{min}\approx
\Gamma_\sigma^3/(4\Omega_\mathrm{L}^2)$. In the intermediate region,
$\Gamma_\mathrm{min}<\Gamma<\Gamma_\sigma$, correlations from the
central peak are bunched (and equal to 3 in the ideal case), due to
the mixture of scattered and emitted light.

In real lasers, $\gamma_\mathrm{L}\neq 0$ due to phase
fluctuations. Non-monochromatic theoretical models, where the phase
varies stochastically~\cite{kimble77b,arnoldus86a}, take into account
both the amplitude and speed of fluctuations, recovering not only a
finite $\gamma_\mathrm{L}$ for the laser
mode~\cite{agarwal76a,kimble77b} but also physical correlations at all
limits~\cite{wodkiewicz80a}. The Jaynes-Cummings model or one-atom
laser is free from this pathology as fluctuations are intrinsic to the
dynamics, provided by the interplay between coherent exchange,
incoherent pump and decay. However, not being fully coherent unless
very far above threshold, one requires a very good system
($\gamma_a\ll g$) in a very intense and well defined lasing regime,
where $\gamma_\mathrm{L}\ll \gamma_a, \Gamma_\mathrm{min}$, that is
$P_\sigma \gg \sqrt{2}g$.

Like for any other system, very broad filters $\Gamma \gg \Omega_+$
recover the full antibunching of the two-level system, therefore
$\lim_{\Gamma\rightarrow \infty}g_\Gamma^{(2)}[\sigma]
(0,0)=g_\sigma^{(2)}[\sigma]=0$. This is independent of the type of
excitation. Under very low coherent excitation (dotted red line in
Fig.~\ref{fig:7}(b)), where the 1PS is dominated by the coherent part,
correlations monotonously go from 1 to 0 as
\begin{equation}
  \label{eq:TueSep18140805CEST2012}
  \lim_{\Omega_\mathrm{L}\rightarrow 0} g_{\Gamma}^{(2)}[\sigma] (0,0)=\Big(\frac{\gamma_\sigma}{\gamma_\sigma+\Gamma}\Big)^2\,.
\end{equation}
This regime has been exploited to create an ultra-coherent single
photon source, i.e., indistinguishable in frequency thanks to the
inherited long laser coherence
time~\cite{gibbs76a,hoffges97a,volz07a,nguyen11a,matthiesen12a}. It is
interesting to note that filtering the coherent peak,
$\Gamma<\gamma_\sigma$, yields the coherent statistics of a laser and
to the destruction of the antibunching even if the coherent part
strongly dominates the 1PS. Only the full emission has the property of
interfering destructively and provide photons one by one.
 
Finally, the impact of detuning from the laser (in the semiclassical
approximation) is shown in Fig.~\ref{fig:8}. In the absence of pure
dephasing, the 1PS remains symmetric around the laser frequency (here
$\omega_\mathrm{L}=0$). Interestingly, the 2PS does not, even if
detuning is small as compared to the driving, as shown in (b). The
asymmetry in the 2PS consists in the disappearance of the ring of
interferences nearer to the two-level system frequency (in this case
$\omega_\sigma >\omega_\mathrm{L}$). When detuning dominates and
$\Omega_+\approx \omega_\sigma$, as in (c), the region around this
peak becomes more antibunched, as it correspond to a real transition
in a two-level system. On the other hand, the region around the other
sideband, $(\Omega_-,\Omega_-)$, becomes more bunched, as it
corresponds to a virtual laser
transition~\cite{cohentannoudji79a,dalibard83a}. Two red lines cross
at $(\Omega_-,\Omega_-)$, only interrupted by the blue interference
ring. A strong bunching point emerges at $(\Omega_+,\Omega_-)$,
accompanied by the typical cascade $\tau$-dynamics, making the system
suitable for heralded single photon emission~\cite{ulhaq12a}.

One can understand the asymmetry in the 2PS with detuning,
$g_\Gamma^{(2)}[\sigma](\Omega_+,\Omega_+)\neq
g_\Gamma^{(2)}[\sigma](\Omega_-,\Omega_-)$, as a manifestation of the
dissimilar two-photon dynamics of the dressed states, namely, the
$\tau$-dynamics of the colour-blind correlations corresponding to each
transition: $g^{(2)}[\ket{-}\bra{+}](\tau)$ and
$g^{(2)}[\ket{+}\bra{-}](\tau)$. These two functions are identical in
the limiting $\tau$-values, equal to 0 at $\tau=0$ and to 1 at
$\tau\rightarrow \infty$, but evolve differently at intermediate
$\tau$. In the limit of high excitation and full filtering of each
peak, such dressed state two-photon dynamics is overall well mapped by
the frequency-resolved correlations,
$g_\Gamma^{(2)}[\sigma](\Omega_+,\Omega_+,\tau)$ and
$g_\Gamma^{(2)}[\sigma](\Omega_-,\Omega_-,\tau)$, as long as $\tau
>1/\Gamma$. However, as we have shown, the $\tau=0$ value is not well
mapped, being larger than zero when the detection is included
(spoiling perfect antibunching). The frequency-resolved functions at
$\tau=0$, include an average over the short-time dressed-state
dynamics, within the detector time-scale, $1/\Gamma$. Therefore, the
interaction with the filter/detector can be considered a dephasing
mechanism of the dressed-state dynamics, which breaks the symmetry of
the 2PS even at $\tau=0$. This view is confirmed by the behaviour of
the physical time-dependent 1PS~\cite{eberly77a}, in the absence of
pure dephasing, which is also asymmetric in general before reaching
its symmetric steady state lineshape (not shown).

\section{Conclusions and further directions}
\label{sec:ThuJun14165142CEST2012}

In conclusion, we have performed the first systematic investigation of
the two-photon spectra (2PS) of a variety of quantum optical systems,
using our recently developed formalism to compute frequency and time
resolved $N$-photon correlations~\cite{delvalle12a}. We have focused
here on the case $N=2$ of two-photon correlations at zero time
delay. Thanks to our formalism, such investigations can be generalised
to higher photon number and/or arbitrary time delays. We studied
systems of increasing complexity, starting from the simplest possible
case that is the quantum harmonic oscillator and proceeding with the
two-level system and then all their possible combinations, namely, two
coupled harmonic oscillators, two two-level systems and a mixture of
both that amounts to the celebrated Jaynes-Cummings model. The latter,
in contrast with our starting point, is an extremely rich and
complicated system that stands as the pillar of cavity quantum
electrodynamics. We have outlined what are the common features shared
by all these systems and what are those specific to their underlying
structure and dynamics. These results constitute the backbone for
two-photon spectroscopy. The main finding can be summarised as
follows:
\begin{itemize}
\item There is a universal bunching of photons when filtering an
  emission line below its linewidth, regardless of its inherent
  statistics, due to the indistinguishable and uncorrelated arrival of
  photons (\emph{indistinguishability bunching}). This manifests as a
  diagonal on all 2PS.
\item An antibunching arises for a two-level system with a
  characteristic butterfly-shape that is reproduced by any two-level
  transition that can be resolved in isolation from a more complex
  level structure (see Ref.~\cite{arXiv_delvalle12c} for an example
  within a four-level system).
\item The 2PS of various types of coupled systems differ greatly from
  each other even when their 1PS are identical, and within the same
  system, some symmetries of the 1PS are lifted in 2PS.  This shows
  the much higher degree of characterization accessible via two-photon
  spectroscopy.
\item The 2PS of coupled harmonic oscillators features an anticrossing
  (two hyperbolas) of uncorrelated emission corresponding to the
  independent and classical polaritons dynamics.  Coupled two-level
  systems or the Jaynes--Cummings model on the other hand give rise to
  rich patterns of correlations permitting to investigate the
  underlying physical picture in terms of relaxation between quantum
  states.
\item Such correlations are due to fundamental processes that can be
  identified through simple equations that locate them in the
  2PS. They are $i)$~\emph{leapfrog processes}, i.e., two-photon
  emission jumping over an intermediate rung of the level structure
  through a virtual state of indeterminate frequency and
  $ii)$~\emph{polariton-to-virtual-state} anticorrelations, where the
  emission does not correspond to dressed states transitions but take
  into account the dynamical nature of the system, with the final
  state provided by a virtual state or fluctuation of the system.
\item Leapfrog processes are particularly noteworthy for applications
  in quantum information processing. They are present in any nonlinear
  quantum system, as simple as an anharmonic oscillator.  $N$-photon
  leapfrogs with $N>2$ also take place, though the higher the $N$, the
  weaker the process.  In the Jaynes--Cummings system, they are
  evidenced by higher order correlations,
  $g_\Gamma^{(N)}(\omega_1,\dots,\omega_N)$, at the frequencies
  \begin{eqnarray}
    \label{eq:WedAug8161036CEST2012}
    \sum_{i=1}^N \omega_i&= N \omega_a \pm (E_n - E_{n-N})\quad \mathrm{or}
    \label{eq:WedAug8173959CEST2012}\\
    &= N \omega_a \pm (E_n + E_{n-N})\,, \label{eq:WedAug8174013CEST2012}
  \end{eqnarray}
  but their further characterization is out of scope of this text.
\item The intricate pattern of $N$-photon correlations that is formed
  by the combination of all the above processes in the
  Jaynes--Cummings model, that is neatly resolved in a system well
  into the strong-coupling regime, evolves when brought in the lasing
  regime into the qualitatively different 2PS of the Mollow triplet,
  that features new types of patterns such as circles rather than
  simply straight lines.  For high rungs of excitation, $n\gg 1$, the
  lines in Eq.~(\ref{eq:WedAug8173959CEST2012}) converge to the
  central antidiagonal, $\sum_i \omega_i\approx 0$, while the lines in
  Eq.~(\ref{eq:WedAug8174013CEST2012}) agglomerate at the outer
  positions, $\sum_i \omega_i\approx \pm 2\sqrt{n}g$. This formation
  of the Mollow triplet in the 1PS gives rise to \emph{leapfrog
    triplet} in the 2PS, with the same origin as in the linear regime
  in terms of two-photon transitions between dressed states.
\end{itemize}

Beyond spelling out the dynamics of emission at the quantum level, the
theory of two-photon spectroscopy also allows to address fundamental
theoretical issues of the quantum formalism and illustrates the deep
link between the quantum dynamics and the detection process. For
instance, in the limit of an ideal detector, the semiclassical
description of the Mollow triplet fails to recover the fundamental
indistinguishability bunching. This is due to the artificial
$\delta$~line of the scattering peak produced by the semiclassical
approximation.  This shortcoming can be solved by turning to a fully
quantized theory or upgrading the semiclassical theory to get rid of
the artifacts caused by a fixed phase.

Another conclusion from our study is that the dynamics of ``real
processes'', involving photon by photon de-excitations, are related to
the system parameters and less influenced by detection. The
corresponding correlations are best resolved when the related peaks
are separated and fully filtered. On the other hand, ``virtual
processes'', involving virtual states such as the leapfrogs and
polariton-to-virtual state interferences, happen within the time of
interaction with the detector, $1/\Gamma$, and conserving energy
within its linewidth, $\Gamma$. Therefore, their correlations become
more prominent using narrow filters.

We have thus amply demonstrated that two-photon spectroscopy unravels
a rich two-photon dynamics and interference effects in open quantum
systems by a precise disposition of filters of given resolutions. Such
correlations can be further taken advantage of by considering finite
time delays and/or optimising the frequency windows. Further
interesting systems to apply two-photon spectroscopy are ultra strong
coupling systems~\cite{ridolfo12a}, closely spaced atoms in optical
lattice~\cite{degenfeldschonburg12a}, the biexciton two-photon
emission in a quantum dot~\cite{delvalle11d,arXiv_delvalle12c} or the
dynamics of Bose-Einstein condensation~\cite{ardizzone12a}.

\section*{Acknowledgements}
  AGT acknowledges support from the FPU program AP2008-00101 (MICINN);
  FPL from the RyC program; CT from MAT2011-22997 (MINECO) and
  S-2009/ESP-1503 (CAM); MJH from the Emmy Noether project HA 5593/1-1
  and from CRC 631 (DFG) \& EdV from the Alexander von Humboldt
  foundation.

\section*{References}

\bibliographystyle{unsrt}
\bibliography{Sci,books,arXiv}
\end{document}